\documentclass[prb,twocolumn,floatfix,showpacs]{revtex4}
\usepackage{psfig}
\begin{document}
\title{{\bf Optical symmetries and anisotropic transport in high-T$_{c}$
superconductors}}
\author{T.P. Devereaux}
\affiliation{Department of Physics, University of Waterloo, Waterloo, ON,
N2L 3G1, Canada}
\date{\today}
\begin{abstract}
A simple symmetry analysis of in-plane and out-of-plane transport in a
family of high temperature superconductors is presented. It is shown that
generalized scaling relations exist between the low frequency electronic Raman response
and the low frequency in-plane and out-of-plane
conductivities in both the normal and superconducting states of the
cuprates. Specifically, for both the normal and superconducting state, the
temperature dependence of the low frequency $B_{1g}$ Raman slope
scales with the $c-$axis conductivity, while
the $B_{2g}$ Raman slope scales with the in-plane conductivity.
Comparison with experiments
in the normal state of
Bi-2212 and Y-123 imply that the nodal transport is largely doping independent and
metallic, while transport near the BZ axes is governed by a quantum critical point
near doping $p\sim 0.22$ holes per CuO$_{2}$ plaquette.
Important differences for La-214 are discussed. It is also
shown that the $c-$ axis conductivity rise for $T\ll T_{c}$ is a consequence of
partial conservation of in-plane momentum for out-of-plane transport.

\end{abstract}
\maketitle

\section{Introduction}
The strong anisotropy of in-plane ($ab$) and out-of-plane ($c$) transport
in the cuprate systems revealed
by angle-resolved photoemission spectroscopy (ARPES), NMR, resistivity, Hall,
Raman, and optical conductivity measurements
is as unresolved and longstanding a problem as superconductivity itself
\cite{ARPES,CooperGray,caxis,Watanabe97,Kendziora93,Forro93,Spectral,Tallon}.
As a function of hole doping per CuO$_{2}$
plaquette $p$ the $ab$-plane resistivity $\rho_{ab}(T)$ (Fig. 1A)
shows a metallic temperature dependence ($d\rho/dt > 0$)
for a wide range of doping
while the $c$-axis resistivity $\rho_{c}(T)$ (Fig. 1B)
varies as $T^{r}$ with an exponent
$r$ that changes from 2 to -2 as $p$ decreases.
The resistivity ratio $\rho_{c}(T)/\rho_{ab}(T)$ is large and
becomes increasingly temperature dependent in all (hole-doped) cuprate systems for
$p$ below $\simeq 0.22$ at low
temperatures\cite{CooperGray,caxis,Watanabe97,Kendziora93,Forro93,Spectral,Tallon}.

\begin{figure}
\psfig{file=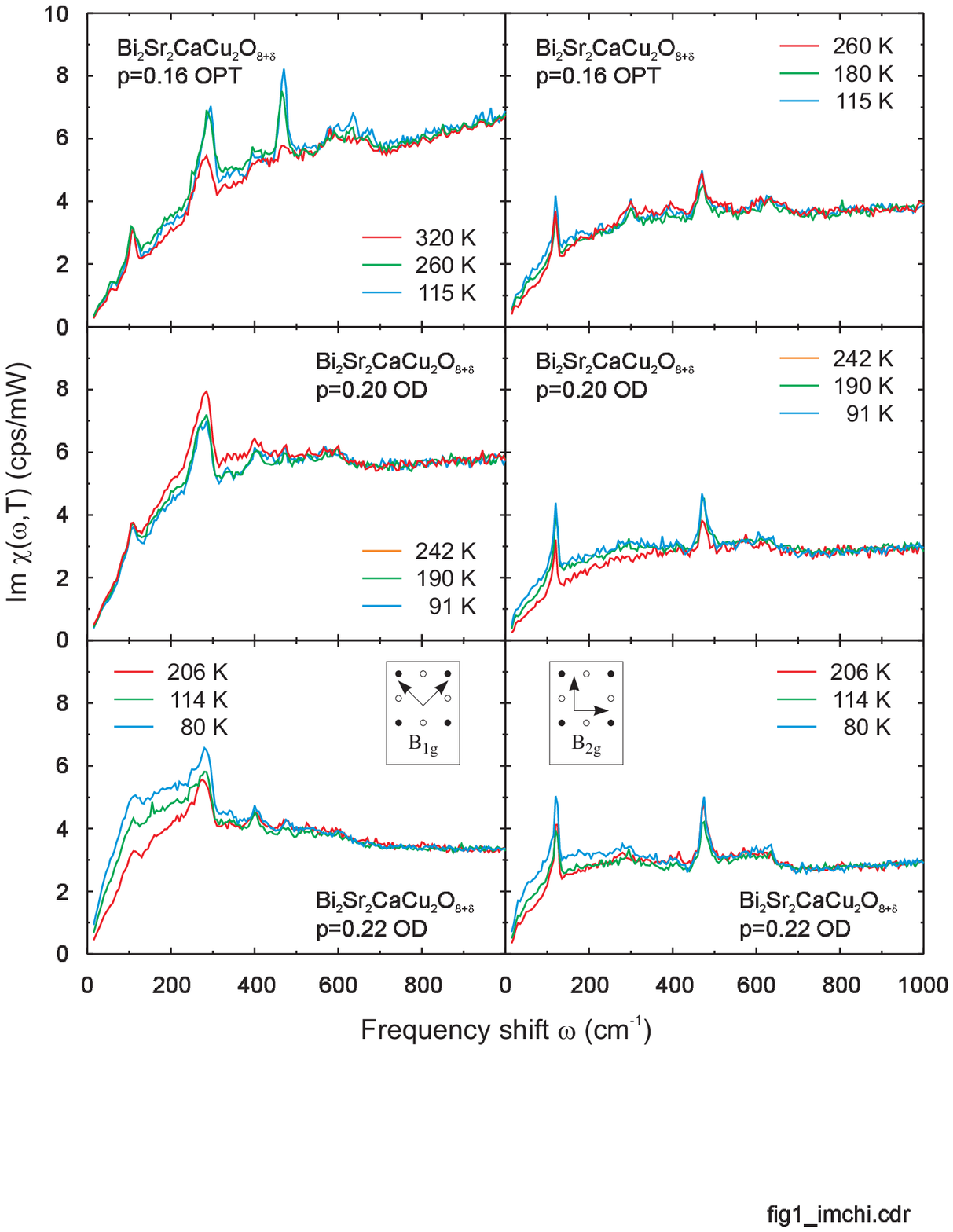,height=7.cm,width=8.5cm,angle=0}
\caption[]{Experimental results for Bi-2212
for $\rho_{ab}(T)$ (Panel A), $\rho_{c}(T)$ (Panel B),
the Raman-derived $B_{2g}, B_{1g}$ qp
relaxation rate $\Gamma^R_{2,1}$ (Panel C, Panel D), respectively.
The solid lines, circles
correspond to underdoped samples ($p=0.10$) with $T_{c}\sim 57$K, dotted lines, squares correspond
to optimally doped samples ($p=0.15$) with $T_{c}\sim 92$K, dashed lines, diamonds correspond
to slightly overdoped samples ($p=0.19$) with $T_{c}\sim 82$K, and the dotted-dashed lines, triangles
correspond to overdoped samples ($p=0.23$) with $T_{c}\sim 52$K. All resistivities were measured in
Ref. \cite{Watanabe97}, except for the overdoped ($T_{c}=52$K) sample which was measured
in Ref. \cite{Kendziora93}. The Raman qp relaxation rates are taken from Ref. \cite{opel2000}.
}\label{fig1}
\end{figure}

It was pointed out early on that the $c-$axis properties provided an useful spectral
tool to examine in-plane charge dynamics\cite{AJL}.
As a result, many approaches have been put forward to address the nature of
in-plane versus out-of-plane transport in terms of anisotropy of the
in-plane quasiparticle (qp)
self energies $\Sigma({\bf k},T)$, $c$-axis hopping $t_{\perp}({\bf k})$,
impurity assisted hopping, interband transitions,
or deconfinement of electrons
\cite{AJL,IoffeMillis,KimCarbotte,Cornaglia01,XiangWheatley,XiangHardy,vanderMarel,cold,hot,Coleman,Varma,abrik,atkin}.
Recently the issue of spectral
weight transfers in optical conductivity measurements brought
about by superconductivity has attracted a great deal of attention\cite{Spectral,Tallon,AJL,IoffeMillis}.
The mechanism by which 3D superconducting phase
coherence sets in is of continued interest and debate which has been guided
in a large part by the measurements of the $c-$axis transport properties.

The issue is still largely unsettled basically due to the open
question of whether electron hopping in the out-of-plane direction
is coherent\cite{Forro93,Spectral,Tallon,AJL,IoffeMillis,KimCarbotte}. If
there were an at least partial conservation of the in-plane
momentum for qp tunnelling along the c-axis, LDA\cite{oka} would
indeed predict an interrelation between c-axis transport and the
qp scattering rate close to $(\pi,0)$ in the Brillouin zone (BZ).
What would be extremely useful would be a transport measurement
beside conductivity which might directly test whether
transport in the plane is intimately tied to out-of-plane
transport.

A behavior similar to the resistivity anisotropy is reflected in
electronic Raman scattering measurements when comparing the
temperature dependence of the low energy continuum measured in
$B_{1g}$ polarization orientations, which project out charge
fluctuations near the BZ axes, to $B_{2g}$ configurations, which
probe charge fluctuations along the BZ diagonals.
Hackl {\it et al.}\cite{Hackl} and
Blumberg and Klein\cite{Blumberg} have
pointed out the close connection between $B_{2g}$ Raman and in the ab-plane
conductivity. Further, Opel {\it et al.}
\cite{opel2000} and Venturini {\it et al.}
\cite{QCP} compared the Raman relaxation rate
in each channel, defined as the inverse of the slope of the low
energy Raman response $\Gamma^{R}_{1,2} = lim_{\Omega\rightarrow
0}[
\partial \chi^{\prime\prime}_{\gamma,\gamma}(\Omega,T)/\partial \Omega]^{-1}$.
For both YBa$_{2}$Cu$_{3}$O$_{7-\delta}$ (Y-123) and
Bi$_{2}$Sr$_{2}$CaCu$_{2}$O$_{8+\delta}$ (Bi-2212), it was found
that for $B_{2g}$ symmetry, $\Gamma^{R}_{2}$ (Fig. 1C)
approximately scales with $\rho_{ab}(T)$ over a wide doping range,
while for $B_{1g}$, $\Gamma^{R}_{1}$ (Fig. 1D) was found to cross
over from metallic to insulator behavior for $p$ less than $\sim
0.22$, occurring a higher dopings than that usually attributed to
the formation of a pseudogap\cite{Tallon}. This has recently been
interpreted as evidence for an underlying quantum critical point
lying near $p_{c}\simeq 0.22$ of an unconventional metal-insulator
transition (MIT)\cite{QCP}.

At low frequencies for underdoped systems, $\sigma_{c}(T)$
for Y-123 and YBa$_{2}$Cu$_{4}$O$_{8}$ (Y-124) decreases
rapidly with decreasing temperature\cite{Homes}. From this a pseudogap has been inferred and
well-documented. A much weaker spectral weight reduction is seen for
La$_{2-x}$Sr$_{x}$CuO$_{4}$ (La-214)\cite{timusk}. Contrary to the out-of-plane conductivity,
it is widely believed that there is no direct indication
of a pseudogap in $\sigma_{ab}$\cite{Puchkov}.
The weak dependence with temperature of the $ab$-plane optical sum
rule compared the rapid decrease at low temperature of the integrated
$c-$axis conductivity has been interpreted in Ref. \cite{IoffeMillis}
qps located near $(\pi,0)$ from the anisotropy of $t_{\perp}$.

Raman scattering has been widely used to address the pseudogap.
Recently, the presence of a pseudogap has been derived from c-axis $A_{1g}$ Raman measurements in
Y-124\cite{Quilty}. A much weaker signature of a pseudogap is seen in the $B_{2g}$ channel
in Y-123 and Bi-2212\cite{opel2000,Nemet}.
In optimal and overdoped systems, pair-breaking features appear only when superconducting
coherence is established and their location
at different energies for different symmetry channels has been well
documented and interpreted as Cooper pairs having $d_{x^{2}-y^{2}}$ symmetry and
well defined low energy qp excitations\cite{Ramanold,tpdapk}.
While the $B_{2g}$ pair-breaking feature appears at and scales with $T_{c}$ for all dopings considered,
closer to optimal doping and for underdoped systems, low frequency spectral weight is lost at low
temperatures and the $B_{1g}$ pair-breaking peak becomes difficult to distinguish from the
background\cite{Ramanold,tpdapk,opel2000,B1gpair,Nemet,Genoud,BlumbergScience,Hewitt,Quilty,footnote}.
This loss of spectral weight with temperature is very similar to the behavior seen in Raman
scattering in Kondo and mixed-valent insulators and is indicative of
gapped excitations\cite{Cooper}.

In the superconducting state, the temperature dependence of the
$ab$-plane low frequency (or regular part of the dc)
conductivity\cite{Hosseini99,Bonn93} typically shows
a peak around 35$K$ which is material dependent and has been
attributed to the rapid collapse of the quasiparticle (qp)
inelastic scattering rate below T$_{c}$ and the rise of the
qp elastic scattering rate for low $T$
\cite{Hirschfeld97,Hirschfeld94,Hensen97,WalkerDuffy}. A similar
peak seen in in-plane thermal conductivity measurements
was found to be sensitive to annealing conditions\cite{Matsukawa96}.
The $c-$axis low frequency
conductivity in YBa$_{2}$Cu$_{3}$O$_{6.95}$ meanwhile does not
show a peak in this region but has an upturn at temperatures below
$25K$. The origin of the peak is currently not
understood\cite{Hosseini98,XiangHardy}. The $c-$axis thermal
conductivity was found to show a very weak peak also sensitive to
annealing conditions\cite{Matsukawa96}. Much less is
known about the temperature dependence of Raman scattering in the
static limit in the superconducting state, although some
theoretical treatments have appeared\cite{tpdapk,WuCarbotte}.
One would like to test whether features shown in
conductivity measurements are found in Raman scattering
measurements and vice-versa.

Shastry and Shraiman have noted the close similarity between the conductivity
and the Raman response and have suggested a scaling relation exists between the two
which follow the same temperature and frequency dependence\cite{ssvr}:
\begin{equation}
\Omega\sigma^{\prime}(\Omega,T)=A\chi^{\prime\prime}(\Omega,T),
\label{SS}
\end{equation}
with $A$ a constant independent of frequency and temperature.
This Shastry-Shraiman (SS) relation holds if the qp self energy
$\Sigma$ is independent of {\bf k} and
has been shown to be exact for both the Falicov-Kimball\cite{tpdjkf1}
and Hubbard\cite{tpdjkf2} models in the limit of large dimensions where the
self energy and vertex corrections are local.
Generally though any {\bf k}-dependence of $\Sigma$ and/or the
irreducible Raman or current vertices
invalidates the SS scaling relation making it inappropriate for
strongly anisotropic systems such as the cuprates.

However, an approximate scaling relation may hold for certain
cases and one purpose of this paper is to point out some of the
connections between the conductivity and Raman response for
strongly anisotropic systems and derive appropriate scaling
relations. In particular we will, based on symmetry arguments,
determine that a variant of the SS relation can be formulated to
show that a scaling relation exists separately between
$\sigma_{ab}$ and $1/\Gamma^{R}_{2}$ and between $\sigma_{c}$ and
$1/\Gamma^{R}_{1}$ as a consequence of the momentum dependence of
$t_{\perp}({\bf k})$, in-plane self energy $\Sigma({\bf k})$, and
a $d_{x^{2}-y^{2}}$ energy gap $\Delta({\bf k})$.
These scaling relations are also found to
also hold in the superconducting state.
Comparison with the available data on Y-123 and B-2212
in the normal state suggests that qps located
near the BZ axes or ``hot spots'' become gapped above optimal
doping\cite{QCP} while the qps located along the BZ diagonals or
``cold spots'' are largely doping independent and remain metallic. 
Thus the c-axis transport is partially influenced by a correlation gap near $(\pi,0)$
because of partial conservation of the in-plane momentum
in c-axis transport and not completely by c-axis diffusion.
There are important differences however with La-214.
Various models for qp scattering as
a function of doping are discussed, and it is found that generally
no single model can adequately capture the complex nature of
electron dynamics over a wide range of doping.
Features of the
theory in the superconducting state qualitatively describe the
behavior seen in the $c-$axis conductivity, but there are
important questions left unanswered.
In conclusion, experimental
evidence in both the normal and superconducting states suggest
that the in-plane momentum is at least partially conserved in
c-axis transport over a very wide doping range.

The plan of the paper is as follows: Sections II and III present
the formalism used and the results for the temperature dependence
of the low frequency in-plane and out-of-plane conductivity and
Raman response in the normal and superconducting states,
respectively, for the common model where $t_{\perp}$ vanishes
along the BZ diagonals, summarized in the Appendix.  The results
are summarized and open points are discussed in Section IV.

\section{Normal State}

\subsection{Formalism}

The quantum chemistry of the tetragonal Cu-O system yields an
out-of-plane hopping which is modulated by the in-plane momentum
in such a way that it is strongly governed by qps located along
the BZ axes as opposed to qps along the zone diagonal,
$t_{\perp}({\bf k}) =t_{\perp}^{0}
[\cos(k_{x}a)-\cos(k_{y}a)]^{2}$, as reviewed in the Appendix.
This
form for the hopping has been widely used to study the penetration
depth\cite{XiangWheatley}, c-axis
conductivity\cite{Spectral,XiangHardy,vanderMarel}, and bi-layer
splitting\cite{oka} in ARPES\cite{ARPES}.
However we note that inclusion of the Cu-O chains or O
displacements would lower the symmetry with the consequence that
the out-of-plane hopping would no longer vanish along the BZ
diagonals which could only be noticeable at very low temperatures.
We now explore the consequences of such a term on the regular part of the
dc conductivities and the
symmetry-dependent electronic Raman response for qp scattering in
the $a-b$ plane and along the $c-$axis in the following sections.

In linear response theory, expressions for the regular part of the
conductivity and Raman response
in the absence of vertex corrections are given as
(here and throughout we set $k_{B}=\hbar=1$)
\begin{eqnarray}
\left(
\begin{array}{c}
\Omega\sigma^{\prime}_{\alpha,\beta}(\Omega)\\
\chi^{\prime\prime}_{\gamma,\gamma}(\Omega)\\
\end{array}
\right)
&&=
\int{dx\over{\pi}}[f(x)-f(x+\Omega)]\\
&&
\times\sum_{\bf k}
\left(
\begin{array}{c}
j^{\alpha}_{\bf k}j^{\beta}_{\bf k}\\
\gamma^{2}_{\bf k}\\
\end{array}
\right)
G^{R}_{\bf k}(x)G^{A}_{\bf k}(x+\Omega).
\label{formal}
\nonumber
\end{eqnarray}
Here $f$ is the Fermi function, $G^{R,A}$ are the retarded, advanced Green's
functions, respectively, $j^{\alpha}_{\bf k}= e{\partial
\epsilon_{\bf k}\over{\partial k_{\alpha}}}$ is the current vertex
for direction $\alpha$ given in terms of the band dispersion
$\epsilon_{\bf k}$ and electron charge $e$, and $\gamma_{\bf k}$
is the Raman vertex set by choosing the incoming and
outgoing light polarization vectors.

The inclusion of vertex corrections is crucial for satisfying Ward
identities for the conductivity and particle number conservation
for the charge density response. They convert scattering lifetimes into 
transport lifetimes, and
also add an additional
source of momentum and temperature dependence to the corresponding
response functions. Vertex corrections have been recently been
considered in FLEX treatments of the Hubbard model\cite{Manske}
and a spin-fermion model\cite{NAFL} where it was shown the
$B_{1g}$ Raman irreducible vertex is highly renormalized near the
$(\pi,0)$ regions of the BZ. In addition vertex corrections have
been calculated exactly in the limit of large dimensions for the
Falicov-Kimball model, where it was shown they are important in
the $A_{1g}$ channel to properly lead to gauge invariance and
particle-number conservation but do not contribute to other
channels\cite{tpdjkf1}. Generally, vertex corrections have not yet
been generically or systematically investigated in 2D and we thus
neglect them since we are interested in exploring simple symmetry
properties of the various experimental probes.

The current vertices are simply $j^{x}_{\bf k}=
v_{x}\sin(k_{x}a)$, and $j^{z}_{\bf k} =v_{z}
[\cos(k_{x}a)-\cos(k_{y}a)]^{2}$, where $v_{x}\sim t$ and
$v_{z}\sim t_{\perp}^{0}$ have only a mild momentum dependence. In
the limit where the incident and scattered photon energies are
small compared to the bandwidth the Raman vertex is given as the
curvature of the band: $\gamma_{\alpha,\beta}= {\partial^{2}
\epsilon({\bf k})\over{\partial k_{\alpha}\partial k_{\beta}}}$
\cite{resonant}. The
vertices are thus determined from the above band structure as
$\gamma_{\bf k} = b_{1} [\cos(k_{x}a)-\cos(k_{y}a)],
b_{2} \sin(k_{x}a)\sin(k_{y}a)$ for $B_{1g}, B_{2g}$
orientations, respectively, while for $c-$axis $A_{1g}$ Raman $\gamma_{\bf
k}=a_{zz}\cos(k_{z}c) [\cos(k_{x}a)-\cos(k_{y}a)]^{2}$. The
prefactors $b_{1}\sim t, b_{2}\sim t^{\prime},$ and $a_{zz}\sim
t_{\perp}^{0}$ can also be assumed to be only mildly frequency
dependent corresponding to
off-resonant scattering and therefore are only multiplicative
constants. Since the energy range considered is very small in
comparison to all electronic bandwidths involved the assumption
$b_{1,2}$ and $a_{zz}$ to be constant is robust under all
realistic circumstances.

As can be seen by the weighting of the vertices,
we may expect similar behavior for the $B_{2g}$ Raman and in-plane
conductivity, and the $B_{1g},$ $c-$axis $A_{1g}$ Raman, and the
out-of-plane conductivity as well. The former two quantities assign weight around the
Fermi surface
(FS) to the diagonals while the latter three assign weight along the zone axes.

In correlated electron systems the density of states (DOS) plays a strong role in determine
transport properties. In Mott insulators, charge transport occurs via
excitations across a Mott gap from the lower to upper Hubbard bands, while in metallic
systems the DOS near the Fermi level plays the dominant role in low
frequency transport. The nature of how the density states evolves across a
MIT has been an issue of intense debate for a large number of years as few exact results are
available. However, in the
limit of large dimensions dynamical mean field theory has a great deal of insight for some model
Hamiltonians\cite{DMFT}. Away from half-filling the Hubbard model and the Falicov-Kimball both
possess metallic ground states. The DOS has a typical three-peak
structure: the separated upper and lower Hubbard bands and a qp DOS at the Fermi
level emerging from the Abrikosov-Suhl resonance in the related impurity problem.
As the system approaches half-filling and/or for larger values of $U$ at fixed filling,
the qp DOS generally diminishes
and vanishes in the Mott insulating phase as spectral weight is transferred into
the Hubbard bands. Capturing this transfer in models in realistic dimensions is one of
the most important and difficult problem in condensed matter physics.

We thus consider charge transport in correlated systems having coherent qps as well as large
energy incoherent charge excitations related to the Hubbard bands. We model coherent qps near
the FS by a phenomenological momentum, frequency, and temperature dependent
self energy derivable in principle from a renormalizable effective low energy theory:
$G^{R,A}_{coh,\bf k}(\omega)= Z_{\bf k}(\omega,
T)/(\omega-\bar\epsilon_{\bf k}\pm i \Gamma_{\bf k}(\omega,T))$.
Here $\bar\epsilon$ is the renormalized band structure,
$Z_{\bf k}(\omega, T)= [1-\partial
\Sigma^{\prime}_{\bf k}(\omega, T)/\partial\omega]^{-1}$ is the qp
residue, and $\Gamma_{\bf k}(\omega,T)$ is the momentum,
frequency, and temperature dependent qp scattering rate. The full Green's function
also includes an incoherent part $G_{inc}$ accounting for larger energy excitations such
as those involving the lower and upper Hubbard bands.
In what follows we focus on low frequency transport in metallic phases
and neglect $G_{inc}$ and singularities of the
self energy indicative of an incipient phase transition.

Converting the momentum sum to an integral over an infinite band
we obtain in the limit of low frequencies
\begin{eqnarray}
&&
\left(
\begin{array}{c}
\sigma^{\prime}_{\alpha,\beta}(\Omega \rightarrow 0,T)\\
\partial \chi^{\prime\prime}_{\gamma,\gamma}(\Omega \rightarrow 0,T)/\partial \Omega\\
\end{array}
\right)
=-2N_{F}\int dx {\partial f(x)\over{\partial x}}\nonumber
\\
&&\times
\Biggl\langle
\left(
\begin{array}{c}
j^{\alpha}_{\bf  k} j^{\beta}_{\bf k} \\
\gamma_{\bf k}^{2} \\
\end{array}
\right)
{Z_{\bf k}^{2}(x,T)
\Gamma_{\bf k}(x,T)\over{\Omega^{2}+
[2\Gamma_{\bf k}(x,T)]^{2}}}\Biggr\rangle,
\label{static}
\end{eqnarray}
where $N_{F}$ is the density of states per spin at the Fermi level and
$\langle \cdots \rangle$ denotes performing an average over
the FS. It can be immediately seen that the SS relation Eq. (\ref{SS})
follows if $\Gamma$ is independent of momentum, as it is in local
theories\cite{DMFT,tpdjkf1,tpdjkf2}.
In what follows we neglect specific features on and off
the FS (such as van Hove) and approximate the 2D FS as a circle
and expand the $c-$axis dispersion for small $t_{\perp}^{0}$ to
obtain:
\begin{eqnarray}
xx ~~\rm{Conductivity}, ~~~~j^{x} = v_{F} \sin(\phi), \nonumber \\
zz ~~\rm{Conductivity}, ~~~~j^{z} = v_{z} \cos^{2}(2\phi), \nonumber\\
B_{1g}~~ \rm{Raman}, ~~~~\gamma_{B_{1g}} = b_{1}\cos(2\phi), \nonumber \\
B_{2g}~~ \rm{Raman}, ~~~~\gamma_{B_{2g}} = b_{2}\sin(2\phi), \nonumber \\
zz~~ A_{1g}~~ \rm{Raman}, ~~~~\gamma_{A_{1g,zz}} = a_{zz}\cos^{2}(2\phi)
\label{simplevertices}
\end{eqnarray}
We note that the $c-$ axis conductivity and
$\partial\chi_{A_{1g,zz}}^{\prime\prime}/\partial \Omega$ are given by the same
expressions, in accordance with the qp scattering rate not having
a $k_{z}$ dependence. Therefore we confirm the SS relation for the
c-axis $A_{1g}$ Raman and c-axis conductivity, respectively:
\begin{equation}
\lim_{\Omega \rightarrow 0}\Omega\sigma_{zz}^{\prime}(T) \propto
\chi^{\prime\prime}_{A_{1g,zz}}(\Omega,T),
\label{a1gzzscaling}
\end{equation}
independent of the form for $\Gamma_{\bf k}$.

At low temperatures we find from Eq. (\ref{static})
\begin{equation}
\left(
\begin{array}{c}
\sigma_{\alpha,\beta}^{\prime}(T)\\
\partial \chi^{\prime\prime}_{\gamma,\gamma}(T)/\partial \Omega\\
\end{array}
\right)
=N_{F}\Biggl\langle
\left(
\begin{array}{c}
j^{\alpha}_{\bf k} j^{\beta}_{\bf k}\\
\gamma^{2}_{\bf k}\\
\end{array}
\right)
{Z_{\bf k}^{2}(T)\over{2\Gamma_{\bf k}(T)}}\Biggr\rangle,
\label{lowT}
\end{equation}
showing the interplay of anisotropies of
the scattering rate and the vertices governing the response
functions.

The simple expressions for $\sigma$ and $\partial\chi^{\prime\prime}/\partial\Omega$
allow for a straightforward
comparison of models for the qp scattering rate. We choose a generic
model which describes strong scattering weighted largely along the BZ
axes plus a temperature dependent scattering rate taken to be uniform around
the FS:
\begin{equation}
\Gamma_{\bf k}(T)=\Gamma_{h}(T)\cos^{2}(2\phi)+\Gamma_{c}(T).
\end{equation}
This form for the qp scattering rate has been widely employed in a
number of models differing in the representations of $\Gamma_{h}$
and $\Gamma_{c}$ constrained only to
possess the full symmetry of the lattice ($A_{1g}$)\cite{cold,hot,Coleman,Varma}.
Further parameterizations of the anisotropy do not lead to appreciable
differences. For the $B_{2g}$ Raman as well as the in-plane
conductivity, the vertices weight out regions of the FS where the
scattering rate is small, along the FS diagonals or ``cold
spots''. However, the $B_{1g}$ and $c-$axis $A_{1g}$ Raman and the
out-of-plane conductivity assign no weight to the diagonals and
thus will be governed by the scattering at the ``hot spots''.

Neglecting the ${\bf k}$-dependence of the qp residue $Z_{\bf k}=Z$, the
resulting integrals can be easily performed to give
\begin{eqnarray}
\sigma_{xx}^{\prime}(T)&=&v_{F}^{2}
{N_{F}Z^{2}\over{2\Gamma_{c}(T)}}{1\over{\sqrt{1+\Gamma_{h}(T)/\Gamma_{c}(T)}}},\\
\sigma_{zz}^{\prime}(T)&=&v_{z}^{2}
{N_{F}Z^{2}\over{2\Gamma_{h}(T)}}
\biggl\{{1\over{2}}-{\Gamma_{c}(T)\over{\Gamma_{h}(T)}}\nonumber\\
&&\times
\left(1-{1\over{\sqrt{1+\Gamma_{h}(T)/\Gamma_{c}(T)}}}\right)\biggr\},
\\
{\partial\chi^{\prime\prime}_{B_{1g}}(T)\over{\partial\Omega}}&=&b_{1}^{2}
{N_{F}Z^{2}\over{2\Gamma_{h}(T)}}\nonumber\\
&&\times\left\{1-{1\over{\sqrt{1+\Gamma_{h}(T)/\Gamma_{c}(T)}}}
\right\},\\
{\partial\chi^{\prime\prime}_{B_{2g}}(T)\over{\partial\Omega}}&=&b_{2}^{2}
{N_{F}Z^{2}\over{2\Gamma_{h}(T)}}{1\over{\sqrt{1+\Gamma_{h}(T)/\Gamma_{c}(T)}}}
\nonumber\\
&&\times\biggl\{1-\sqrt{1+\Gamma_{h}(T)/\Gamma_{c}(T)}+\nonumber\\
&&\Gamma_{h}(T)/\Gamma_{c}(T)\biggr\}.
\end{eqnarray}
These results for the ab-plane and c-axis conductivity have been
derived several times, most recently by Refs.
\cite{XiangHardy,vanderMarel}. However here it can be seen that there is
a direct connection between conductivities and Raman response
functions. It is clear that the
function form for the scattering rate determines the temperature
dependence of all four response functions, and that the SS relation
Eq. (\ref{SS}) does not hold in general.

Early on, ARPES measurements yielded $\Gamma_{c}\ll \Gamma_{h}$ from smeared
spectral functions seen near the BZ axes compared to the BZ diagonals\cite{ARPES}.
However, recent ARPES measurements
indicated bi-layer splitting may have led to an
overestimation of $\Gamma_{h}$\cite{ARPES,ARPES2},
but still the limit $\Gamma_{c}\ll \Gamma_{h}$
is a useful limit to explore. In this limit the response functions are
\begin{eqnarray}
\label{normal1}
&&\sigma_{xx}^{\prime}(T)=v_{F}^{2}
{N_{F}Z^{2}\over{2\sqrt{\Gamma_{c}(T)\Gamma_{h}(T)}}},\\
\label{normal2}
&&\sigma_{zz}^{\prime}(T)=v_{z}^{2}
{N_{F}Z^{2}\over{2\Gamma_{h}(T)}},\\
\label{normal3}
&&{\partial\chi^{\prime\prime}_{B_{1g}}(T)\over{\partial\Omega}}=b_{1}^{2}
{N_{F}Z^{2}\over{2\Gamma_{h}(T)}},\\
\label{normal4}
&&{\partial\chi^{\prime\prime}_{B_{2g}}(T)\over{\partial\Omega}}=b_{2}^{2}
{N_{F}Z^{2}\over{2\sqrt{\Gamma_{h}(T)\Gamma_{c}(T)}}}.
\end{eqnarray}
This directly shows the similarity between the $B_{1g}$ Raman slope and the
$c-$axis conductivity, and $B_{2g}$ Raman slope and the in-plane conductivity, regardless
of the functional form chosen for two contributions to the qp scattering rate. Thus in
this model consistent with experiments, a variant of
the SS relation for the cuprates may be expressed as
\begin{eqnarray}
\lim_{\Omega \rightarrow 0}\Omega\sigma_{xx}^{\prime}(T) \propto
\chi^{\prime\prime}_{B_{2g}}(\Omega,T), \nonumber \\
\lim_{\Omega \rightarrow 0}\Omega\sigma_{zz}^{\prime}(T) \propto
\chi^{\prime\prime}_{B_{1g}}(\Omega,T).
\label{genSS}
\end{eqnarray}
This demonstrates how out-of-plane transport can be directly
inferred from in-plane optical transport measurements. Further,
this confirms the behavior shown in Fig. (\ref{fig1}), indicating
that the in-plane momentum must be at least partially conserved
for transport perpendicular to the CuO$_{2}$ planes. Moreover,
with Eq. (\ref{a1gzzscaling}) this indicates that the c-axis
$A_{1g}$ Raman should scale with $B_{1g}$ Raman:
\begin{equation}
\lim_{\Omega \rightarrow 0}
\chi^{\prime\prime}_{A_{1g,zz}}(\Omega,T)
\propto
\chi^{\prime\prime}_{B_{1g}}(\Omega,T).
\label{a1gb1gscaling}
\end{equation}
Eqs. (\ref{genSS}-\ref{a1gb1gscaling}) are the central results of this section.

When and how might the scaling relations Eqs.
(\ref{genSS}-\ref{a1gb1gscaling}) breakdown? Clearly these scaling relations result from
the momentum dependence of the respective response vertices, and since they are dictated
solely on symmetry grounds, changes in how one represents the momentum dependence of the
vertices can only lead to qualitative effects. However, there are a number of important
factors to consider.
First, the inclusion of $G_{inc}$ will change the scaling relations if there is appreciable
spectral weight near the FS, but if we restrict ourselves
to metallic systems and low frequencies, then these changes are expected to be small.
They might however be large for a system lying near a quantum critical point and the scaling
relations may be violated.
Next, relating the c-axis conductivity to
the $A_{1g}$ c-axis and $B_{1g}$ Raman requires that
the c-axis coherent hopping vanishes along the BZ diagonals. Deviations would come from
incoherent diffusive hopping, or more
complex coherent
hopping paths such as via the Cu-O chains in Y-123, and would result in a mixing in
the scaling properties for in-plane conductivity and $B_{2g}$ Raman transport. Lastly,
vertex corrections can appreciably alter the scaling relations. Vertex corrections yield transport
scattering rates in place of $2\Sigma^{\prime\prime}$ as required by Ward identities, and yield
a $f-$sum rule for the integrated conductivity proportional to minus the kinetic energy.
Ward identities can be useful for the conductivity to show
that vertex corrections vanish for a momentum-dependent self energy, but no Ward identities exist
for Raman with crossed polarization vectors.
For example, vertex corrections may renormalize even-parity momentum charge vertices (Raman) but not
odd-parity current vertices (conductivity). If these scaling relations are found to
hold, they would imply that vertex corrections at low frequencies and c-axis hopping along
the BZ diagonals would play only a very minor role in determining low frequency transport.

\subsection{Transport models}

The scaling relations of Eq.(\ref{genSS}) can be seen from Fig. \ref{fig1} to be qualitatively obeyed.
We now consider several models for $\Gamma_{h}(T)$ and $\Gamma_{c}(T)$ to explore the scaling
relations Eqs.(\ref{normal1}-\ref{normal4}) to address the role of anisotropic qp scattering.
In all models, $\Gamma_{h}(T)$ and $\Gamma_{c}(T)$ are generally
constrained by the estimated width of the spectral function
measured in ARPES experiments\cite{ARPES}. In both
a ``cold spot''\cite{cold} and "hot spot" model\cite{hot}, $\Gamma_{c}(T)$
describes weakly renormalized qp scattering primarily along the FS
diagonals generally of the form
\begin{equation}
\Gamma_{c}(T)=\Gamma_{imp}+ T^{2}/T_{0},
\end{equation}
where $\Gamma_{imp}$ represents elastic impurity scattering and $T_{0}$
is the energy scale of a renormalized Fermi liquid. The
impurity scattering may
be chosen to reproduce the extrapolated $T=0$ resistivity and $T_{0}$
is a parameter to be chosen to fit a cross-over from $T^{2}$ to $T$ in
the resistivity. In the ``hot spot'' model\cite{hot,Chubukov}, $\Gamma_{h}(T)=\sqrt{\Gamma_{hs} T}$ represents
scattering with exchange of antiferromagnetic reciprocal lattice momentum ${\bf Q}$ which
has been widely employed to determine the optical conductivity,
in-plane and Hall resistivity in relation to ARPES. However similar
behavior is also obtained for scattering in systems lying near a charge
ordering instability\cite{DiCastro} or near a FS Pomeranchuk instability\cite{Pomeranchuk1}.
In the ``cold spot'' model\cite{cold},
$\Gamma_{h}(T)$ is taken to be a constant $\Gamma_{hs}$ presumed
to arise from strong $d_{x^{2}-y^{2}}$ pairing fluctuations, and has been
employed in several works
to describe in-plane and out-of-plane optical conductivity and
magneto-transport\cite{IoffeMillis,cold,vanderMarel,Coleman}. However the microscopic origin of
$\Gamma_{hs}$ is unclear in this model.
In the marginal Fermi liquid (MFL) model most recently described in Ref. \cite{Varma},
$\Gamma_{c}(T) \sim T$ and $\Gamma_{h}(T)\sim$ constant due to
impurity scattering in correlated systems whereby strong correlation nearby a point-like
scatterer induce real-space extensions of the impurity potential\cite{extended}.

Following Ref. \cite{opel2000}, the ``Raman scattering rate''
$\Gamma^{R}_{\mu}(T)$ for each channel is defined as the inverse of the Raman slope
$\Gamma^{R}_{\mu}(T)=
\left[{\partial\chi_{\mu}^{\prime\prime}(\Omega\rightarrow 0,T)\over{
\partial \Omega}}\right]^{-1}$
in order to obtain information on the single particle scattering rate on
regions of the FS selected by polarization orientations $\mu=1,2$
for $B_{1g,2g}$, respectively. In the "hot
spot model" we obtain
$\Gamma^{R}_{1}\sim T^{1/2}$ and
$\Gamma^{R}_{2}\sim T^{5/4}$, respectively,
while in the "cold spot" model we obtain
$\Gamma^{R}_{1}\sim$ constant and
$\Gamma^{R}_{2}\sim T$, respectively.
The MFL model yields $\Gamma^{R}_{1}\sim$ constant and
$\Gamma^{R}_{2}\sim \sqrt{T}$, respectively.
None of the models considered have presented analytic forms for the scattering
rate as a function of doping, and presumably in all models
$\Gamma_{h}$ would be expected to be small in overdoped systems.

It is often useful to look at the ``scattering ratio''
$\Gamma^{R}_{1}(T)/\Gamma^{R}_{2}(T)\sim\rho_{c}(T)/\rho_{ab}(T)
\sim T^{-m}$. The models discussed give $m=1/2, 3/4,$ and $1$ for
MFL, ``hot'', and ``cold'' spot models, respectively. These
preceding exponents are summarized in Table I. As can be seen from Fig. (\ref{fig1}),
all of these models can {\it qualitatively} describe the experimental results for
overdoped systems, but important deviations occur for optimal and underdoped systems.
The ``hot spot'' model yields a stronger temperature dependence however than that seen
for the $B_{1g}$ Raman and c-axis conductivity.

\begin{table}
\caption{Summary of the low temperature dependence of
the inverse conductivities, the Raman relaxation rates
$\Gamma^{R}_{\mu}$ and the scattering ratio defined in the text.\label{table1}}
\begin{ruledtabular}
\begin{tabular}{|l|l|l|l|}
Response & MFL & ``Hot spot'' & ``Cold spot''\\
\hline
$\Gamma^{R}_{2}(T), \sigma^{-1}_{xx}(T)$& $T^{1/2}$ & $T^{5/4}$& $T$\\
\hline
$\Gamma^{R}_{1}(T), \sigma^{-1}_{zz}(T)$& constant & $T^{1/2}$& constant\\
\hline $\Gamma^{R}_{1}/\Gamma^{R}_{2},
\sigma_{xx}/\sigma_{zz}$ & $T^{-1/2}$ &
$T^{-3/4}$ & $T^{-1}$\\
\end{tabular}
\end{ruledtabular}
\end{table}

\subsection{Pseudogap}

The upturn of both $\Gamma_1(T)$ and $\rho_{c}(T)$
at low temperatures for optimal and underdoped systems is indicative of gapped qps
and connected to an anisotropic pseudogap largest near the BZ axis\cite{Tallon}.
A major issue\cite{Iyengar} is whether the pseudogap is caused by pairing without long-range
phase coherence or due to loss of well defined qps at the FS
related to the formation of a precursor Mott gap,
or spin-density and/or charge-density wave states, for  example.

In the former case, the superconducting gap amplitude closes at T$^{*}$ while strong phase fluctuations
force the superfluid density to appear at T$_{c}$\cite{nem}. In more exotic phases emerging from
$Z_{2}$ gauge theories, electrons fractionalize away from the BZ diagonals, spinons become deconfined
and holons condense and become gapped\cite{Wen}. In these scenarios one might expect a feature
in the spectra appearing at a high energies which merges into the superconducting feature at T$_{c}$. It is not
immediately clear whether this occurs in Raman data due to the nature of the 600 cm$^{-1}$
peak\cite{footnote}.

In the spin and/or charge precursor scenario, anisotropic SDW and/or CDW fluctuations strongly affect the integrity of qp
excitations near the BZ axes\cite{DiCastro,Chubukov}. Relatedly strong electron and Umklapp scattering in particular
due to the nearness of a nesting condition can drive FS topological
changes near the BZ axes or "hot spots" which preserve\cite{patch} or lower\cite{Pomeranchuk2} the symmetry
of the FS.

It is clear that the pseudogap is a manifestation of strong correlations regardless of which scenario is considered.
Thus we take a simple approach and relate the pseudogap to a correlation gap as a precursor to the Mott insulating
phase regardless of any underlying order, characterized by the development of $G_{inc}$ involving large energy transfers across
a precursor Mott gap. The gapping can thus be crudely understood as the loss of well defined qps located near the $(\pi,0)$
regions of the FS, implying that the coherent part of the Green's function diminishes away from the BZ diagonal.

Therefore in what follows the role of anisotropy in the qp residue $Z$ is explored in a simple effort to model
the effect of a loss of qp transport for the ``hot'' qps with decreasing $p$.
Taking $Z_{\bf k}(T)=Z_{h} e^{-(E_{g}/T)\cos^{2}(2\Phi)}$ as a
phenomenological model of angular dependent gapping of qps with an
energy scale $E_{g}$, the integrals are straightforward and the result can be expressed
analytically in terms of a degenerate hypergeometric function of two variables:
\begin{eqnarray}
&&\begin{array}{l}
\sigma^{\prime}_{xx}(T),\partial\chi^{\prime\prime}_{B_{2g}}(T)/\partial\Omega\\
\sigma^{\prime}_{zz}(T),\partial\chi^{\prime\prime}_{B_{1g}}(T)/\partial\Omega\\
\end{array}
\sim \\
&&{N_{F}\over{2\Gamma_{c}(T)}}
\cases{\Phi_{1}\left({1\over{2}},1,2,-{\Gamma_{h}\over{\Gamma_{c}(T)}},-{E_{g}\over{T}}\right),\cr
\Phi_{1}\left({3\over{2}},1,2,-{\Gamma_{h}\over{\Gamma_{c}(T)}},-{E_{g}\over{T}}\right).}
\nonumber
\end{eqnarray}
For almost all temperatures, the function can be accurately described
as the previous results Eqs. (\ref{normal1}-\ref{normal4}) with the
sole exception that $\sigma^{\prime}_{zz}(T)$ Eq. (\ref{normal2}) and
$\partial\chi^{\prime\prime}_{B_{1g}}(T)/\partial\Omega$ Eq. (\ref{normal4})
are multiplied
by $e^{-2E_{g}/T}$. Thus we note that if qps located near the BZ axis become gapped or lose
their spectral weight at the Fermi level, the $B_{1g}$ Raman slope and $c-$axis conductivity
will show activated behavior while the $B_{2g}$ Raman slope and the in-plane conductivity would
continue to show metallic behavior. This is qualitatively the situation found for doping levels
below $p_{c}\sim 0.22$ in all the cuprates.

\subsection{Comparison with experiments}

\begin{figure}
\psfig{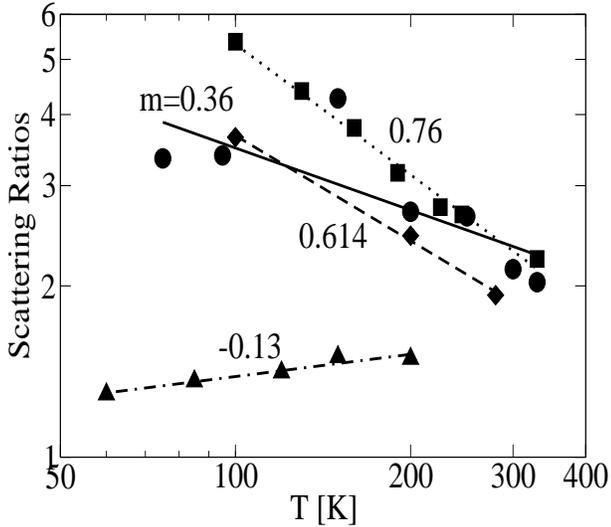}
\caption[]{A log-log plot of the Raman derived ``scattering ratios''
$\Gamma^{R}_{1}/\Gamma^{R}_{2}$ (defined in the text) for Bi-2212
in Ref. \cite{opel2000}
for underdoped (circles, $m=0.36$), optimally
doped (squares, $m=0.76$), slightly overdoped (diamonds, $m=0.614$) and
appreciably overdoped (triangles, $m=-0.13$)
samples shown in Fig. \ref{fig1}, respectively.
The exponent $m$ is determined from a least-squares fit to $T^{-m}$.}
\label{fig2}
\end{figure}

\begin{figure}
\psfig{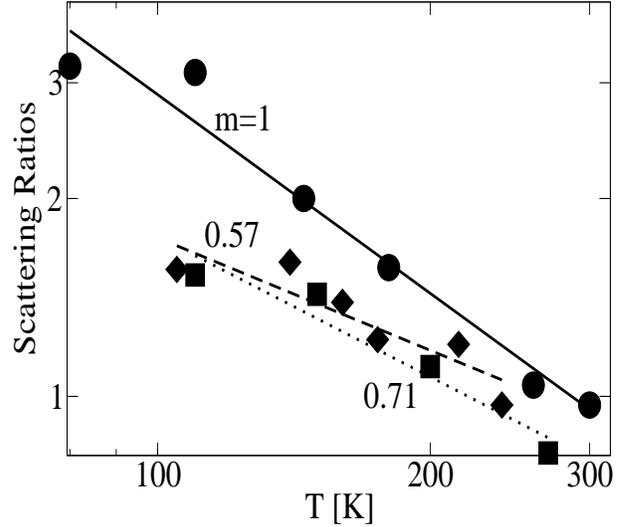}
\caption[]{A log-log plot of the data obtained in Ref. \cite{opel2000} for the
Raman derived ``scattering ratios''
$\Gamma^{R}_{1}/\Gamma^{R}_{2}$ (defined in the text) for
YBa$_{2}$Cu$_{3}$O$_{6.5}$ (circles, $m=1$),
YBa$_{2}$Cu$_{3}$O$_{6.93}$ (squares, $m=0.71$), and
YBa$_{2}$Cu$_{3}$O$_{6.98}$ (diamonds, $m=0.57$), respectively.}
\label{fig3}
\end{figure}

The data for the Raman derived scattering ratio for Bi-2212 is
shown in Fig. \ref{fig2}. The data are derived from the
measurements shown in Fig. \ref{fig1}. The ratio derived from the
measurements on three differently doped samples of Y-123 are shown
in  Fig. \ref{fig3}. For Bi-2212 the ratio slightly increases
($m<0$) with temperature for appreciably overdoped systems, in
agreement with the results obtained for
La-214\cite{Hackl2002}. For
decreasing doping $p$ in both Bi-2212 and Y-123, the exponent $m$
is positive and increases as the ``hot'' qps become gapped and the
``cold'' qps do not appreciably change. The large variation of the
data from the underdoped Bi-2212 sample is due to the small
intensity at low frequencies from which the slope is derived.
Apart from this sample however a power-law fit adequately
describes the data for both compounds.
Near optimum doping both the MFL and ``cold spot'' model
give reasonable agreement for the ``scattering ratio'' while on
the underdoped side the ``cold spot'' model gives an exponent of
$1$ in agreement with the data on Y-123 and in rough agreement
with the data on Bi-2212. An exponential dependence
on temperature has been used in Refs. \cite{Watanabe97} for the
resistivity ratio to determine the magnitude of a pseudogap, for example.
We note that the Raman measurements are not yet of sufficient precision to determine $E_{g}$
from a fit since a straight line fit works well as shown in Fig. \ref{fig2}. The curvature
may be obscured by the small signals measured at low frequencies
however. More accurate data would be very useful.

Recent ARPES measurements have revealed that the qp self energy may not be as anisotropic as determined earlier
due to the more accurate detection of bilayer splitting near the BZ axes\cite{ARPES2}. In addition,
a more quantitative investigation of the qp self energy derived from recent ARPES measurements on overdoped and optimally
doped Bi-2212\cite{ARPES3} has been used
to argue that agreement with the magnitude and temperature dependence of in-plane resistivity measurements on
similar compounds can only be obtained if the transport scattering rate has no contributions from $\Gamma_{hs}$
and is given solely by an MFT dependence\cite{Varma} on all regions of the BZ. 
A similar conclusion has been reached regarding the self energy and optical conductivity\cite{MillisDrew}. 
It is however important to note that the magnitude of the derived resistivity agrees with experiment to
within a factor less than two for temperatures between 100-300K. 
A more or less isotropic qp self energy cannot be reconciled with the Raman data unless vertex corrections
are brought into play. If the experiments are taken at face value, this would imply that vertex corrections affect
the Raman response and the conductivity in opposite ways:
Raman vertex corrections would act to {\it enhance} qp scattering near
the BZ axes while conductivity corrections would {\it decrease} it. This has the important consequence from the
symmetry arguments presented above that the $c-$axis
conductivity should then be more coherent in contrast to experiments. This lack of consistency between ARPES, 
$ab$- and $c$- axes conductivity, and $B_{1g,2g}$ Raman must then be traced to the detailed role of vertex corrections for
each response. This work is currently in progress.

It is important to point out that the results obtained on La-214
are qualitatively different from Y-123 and Bi-2212 in
underdoped systems\cite{Hackl2002}. For
La$_{1.9}$Sr$_{0.1}$CuO$_{4}$,
a clear Fermi liquid like peak
develops at low frequencies {\it in the} $B_{1g}$ {\it channel}
which sharpens as temperature is lowered so that
$\partial\chi^{\prime\prime}(T)/d\Omega$ falls with decreasing
temperature, similar to the behavior of {\it the} $B_{2g}$ {\it
channel} in Y-123 and Bi-2212.
These features appear more or less continuously with
doping. However, the peak in the $B_{2g}$
channel seems to mimic the $B_{2g}$ response in Y-123 and
Bi-2212. We note that this is consistent with ARPES in which a more
smeared spectral function is seen for $(\pi/2,\pi/2)$ rather than
$(\pi,0)$ crossings\cite{Ino}.
Recently strong far infrared peaks have been observed in ab-plane
optical response\cite{Lucarelli} in La$_{2-x}$Sr$_{x}$CuO$_{4}$
for $x=0.05-0.19$ which follow a dependence on $x$
consistent with a coexistence of charge stripes and antiferromagnetic domains\cite{nem}.
Similar strong far infrared peaks have also been observed in Bi$_{2}$Sr$_{2}$CuO$_{6}$
(Bi-2201)\cite{Lupi} and interpreted\cite{DiCastro} in terms of instabilities of a Fermi liquid
to charge ordering. While this interpretation is still open to questions,
both of these observations can be reconciled with Raman scattering
measurements if the stripes were aligned solely along the Cu-O bond
directions. Whether the stripes are conducting or insulating, and whether they are
static or dynamically fluctuating,
the $B_{2g}$ Raman response would have a polarization component
perpendicular to the stripes and thus would project onto incoherent qp transport channels
while the $B_{1g}$ would have a finite project of both the
incident and scattering polarization light vectors along a sector of coherent, conducting excitations
consistent with observations. These simple
symmetry considerations would change if the stripes were thought to be fluctuating in various different
orientations or rotated by 45 degrees, as evidence suggest they might for more underdoped samples.
More data and further calculations are essentially needed to clarify this point. It is an important and
open issue to understand why this occurs for a wide range of doping in La-214 and not Y-123 and Bi-2212.

We note that only limited experimental information exists concerning c-axis Raman
measurements due to the surface problems, but recently Quilty {\it et al.} have shown
that the low frequency c-axis Raman spectral weight
in YBa$_{2}$Cu$_{4}$O$_{8}$ depletes as temperatures are lowered\cite{Quilty}.
In conjunction with
the spectral weight depletion at low temperatures
seen in $B_{1g}$ measurements on the same compound\cite{Genoud},
the admittedly limited experimental evidence is also consistent
with $A_{1g,zz}$ and $B_{1g}$ scaling. More data would of course be useful to check
this further. In this regard it should be mentioned that there is recent evidence that
the c-axis Raman may shed light on a Raman active c-axis plasmon\cite{Munzar}. It would be
extremely useful to examine whether the plasmon would violate the scaling relation or could
possibly lead to a mode-coupling which pushes the plasmon into the $B_{1g}$ channel.

\section{Superconducting state}

\subsection{Formalism}

We now consider how anisotropic transport in the normal state may be reflected in
the superconducting state. In particular we would like to address whether
the variant of the SS relation presented in Eq. (\ref{genSS}) holds in the superconducting state.

In the absence of vertex corrections, the expressions for the Raman
response and the optical conductivity in the static limit are given in terms of
the Nambu Green's functions as:
\begin{eqnarray}
&&\left(
\begin{array}{c}
\sigma_{\alpha,\beta}^{\prime}(T)\\
\partial \chi^{\prime\prime}_{\gamma,\gamma}(T)/\partial \Omega\\
\end{array}
\right)
=2\sum_{\bf k}
\left(
\begin{array}{c}
j^{\alpha}_{\bf  k} j^{\beta}_{\bf k} \\
\gamma_{\bf k}^{2} \\
\end{array}
\right)
\label{superconducting}
\\
&&\times\int {dx\over{\pi}} {\partial f(x)\over{\partial x}}
\left\{G_{0}^{\prime\prime}({\bf k},x)^{2}+
G_{3}^{\prime\prime}({\bf k},x)^{2}\pm
G_{1}^{\prime\prime}({\bf k},x)^{2}\right\}.
\nonumber
\end{eqnarray}
Here
$\hat G({\bf k},\omega)={1\over{\tilde\omega\hat\tau_{0}-\tilde\epsilon({\bf k})
\hat\tau_{3}-\tilde\Delta({\bf k})\hat\tau_{1}}}=
G_{0}({\bf k},\omega)\hat\tau_{0}+
G_{1}({\bf k},\omega)\hat\tau_{1}+
G_{3}({\bf k},\omega)\hat\tau_{3}
$ with the renormalized
quantities determined from the Pauli components of the self energy as
$\tilde\omega=\omega-\Sigma_{0}({\bf k},\tilde\omega),
\tilde\epsilon({\bf k})=\epsilon({\bf k})+\Sigma_{3}({\bf k},\tilde\omega),$
and $\tilde\Delta({\bf k})=\Delta({\bf k})+\Sigma_{1}({\bf k},\tilde\omega)$.

It is well know that vertex corrections appreciably alter
universal results and the Wiedemann-Franz law for $d-$wave
superconductors\cite{DurstLee,tpdapk2000,Chiao2000}. In addition,
they are crucially important for describing the back-flow needed
to restore gauge-invariance in the superconducting state and
appreciably alter the fully-symmetric $A_{1g}$ response over a
wide range of frequencies\cite{dvz}. Again we neglect them to exploit simple
symmetry considerations.
Therefore we only consider $\sigma_{xx}, \sigma_{zz}$ and the
$B_{1g}$ and $B_{2g}$ Raman response. The reader is referred to
Refs. \cite{dvz,DurstLee,tpdapk2000,Chiao2000} where these issues
have been addressed at length.

The self energy is usually broken into an inelastic term,
such as due to phonons or spin-fluctuations, and
an elastic term due to scattering from impurities:
$\hat\Sigma=\hat\Sigma^{inelastic}+\hat\Sigma^{elastic}$\cite{Tmatrix}.
Since the integrand in Eq. (\ref{superconducting})
is weighted out for small frequencies and
since $\Sigma_{1,3}^{\prime\prime}({\bf k},\tilde\omega)$ coming from
inelastic scattering are odd functions
of frequency while $\Sigma_{0}^{\prime\prime}({\bf k},\tilde\omega)$ is even,
we only retain $\Sigma_{0}$. If one considers $s-$wave impurity scattering
in the Born or unitary limit, then $\Sigma_{1,3}^{elastic}$ can be neglected
as well. However generally in other limits and in particular if the impurity
potential is anisotropic as it should be in correlated systems, one must
keep these terms as well\cite{DurstLee,tpdapk2000,Tmatrix}.

In the next subsection the role of disorder in determining the
asymptotic low temperature limit of the results functions is considered,
and then in the
following subsection, inelastic scattering from spin fluctuations is used
to determine the full temperature dependence below $T_{c}$.

\subsection{Disorder}

We first consider scattering from point-like impurities to determine the
low temperature limit of the response functions in the superconducting state.
For $s-$wave impurity scattering
$\tilde\omega=\omega-\Gamma {\bar g_{0}\over{c^{2}-\bar g_{0}^{2}}}$,
with
$\bar g_{0} = {1\over{i}}\langle {\tilde\omega\over{\tilde\omega^{2}-
\Delta^{2}({\bf k})}}\rangle$,
$\Gamma={n_{i}\over{\pi N_{F}}}$, $n_{i}$ the density of impurities,
and $c$ the phase shift\cite{Hirschfeld97}.
The self
energy is determined self-consistently for temperatures below $T^{*}
\sim n_{i}$ due to the formation of a bound-state impurity band at the
Fermi level. In this limit,
the solution may be expanded for small frequencies as
$\tilde\omega=a\omega+i\gamma_{0}+ib\omega^{2}$, with
$a, b$ and $\gamma_{0}$ determined from the impurity concentration
and magnitude of the phase shift\cite{Hirschfeld94}.
Performing the standard integrals in Eq. (\ref{superconducting}) yields
in the limit of low temperatures $T\ll T^{*}$
\begin{eqnarray}
&&\left(
\begin{array}{c}
\sigma_{\alpha,\beta}^{\prime}(T\ll T^{*})\\
\partial \chi^{\prime\prime}_{\gamma,\gamma}(T\ll T^{*})/\partial \Omega\\
\end{array}
\right)
=-N_{F}\int dx \frac{\partial f(x)}{\partial x}\nonumber\\
&&\times\biggl\{
\gamma^{2}_{0}I_{3/2,0}^{\chi_{\gamma,\gamma},\sigma_{\alpha,\beta}}+
x^{2}\biggl[2b\gamma_{0}
I_{3/2,0}^{\chi_{\gamma,\gamma},\sigma_{\alpha,\beta}}+
I_{5/2,0}^{\chi_{\gamma,\gamma},\sigma_{\alpha,\beta}}\nonumber\\
&&\times\left(\frac{15}{2}a^{2}\gamma^{2}_{0}-3b\gamma^{3}_{0}\right)
-\frac{15}{2}a^{2}\gamma^{4}_{0}I_{7/2,0}^{\chi,\sigma}\nonumber\\
&&-\frac{5}{2}a^{2}\gamma^{2}_{0}
c^{\chi,\sigma}I_{7/2,1}^{\chi_{\gamma,\gamma},\sigma_{\alpha,\beta}}\biggr]\biggr\},
\label{mainsuper1}
\end{eqnarray}
with the functions
\begin{eqnarray}
&&I_{\nu,\mu}^{\chi_{\gamma,\gamma}}=
\biggl\langle\frac{\gamma^{2}({\bf k})\Delta^{\mu}({\bf k})}
{[\gamma^{2}+\Delta^{2}({\bf k})]^{\nu}}\biggr\rangle,\nonumber\\
&&I_{\nu,\mu}^{\sigma_{\alpha,\beta}}=\biggl\langle\frac{v_{\alpha}({\bf k})
v_{\beta}({\bf k})\Delta^{\mu}({\bf k})}
{[\gamma^{2}+\Delta^{2}({\bf k})]^{\nu}}\biggr\rangle,
\label{mainsuper2}
\end{eqnarray}
and the constant $c^{\chi,\sigma}=2,0$ for the Raman response and
conductivity, respectively due to the different coherence factors.
Eqs. (\ref{mainsuper1} - \ref{mainsuper2}) reduce to those found
in Ref. \cite{Graf96} for the case of the $a-b$ plane
conductivity. The functions $I_{\nu,\mu}^{\chi,\sigma}$ are
straightforward to compute for a cylindrical FS and $\Delta({\bf
k})=\Delta_{0} \cos(2\phi)$.  For resonant impurity scattering
($c=0$), $a=1/2, b=-\frac{1}{8\gamma_{0}}$, and $\gamma_{0}$ is
determined self-consistently via
$\gamma_{0}=\sqrt{\pi\Gamma\Delta_{0}\over{2\ln(4\Delta_{0}/\gamma_{0})}}$\cite{Hirschfeld94}.
Eqs. (\ref{mainsuper1}-\ref{mainsuper2}) then yield:
\begin{eqnarray}
\sigma^{\prime}_{xx}(T\ll T^{*})={ne^{2}\over{m\pi\Delta_{0}}}
\left[1+\frac{\pi^{2}}{12}\frac{T^{2}}{\gamma_{0}^{2}}\right],
\label{oresults2a}
\\
\sigma^{\prime}_{zz}(T\ll T^{*})={ne^{2}\over{m\pi\Delta_{0}}}
2\frac{v_{z}^{2}}{v_{F}^{2}}
\frac{\gamma_{0}^{2}}{\Delta_{0}^{2}}\left[1-\frac{\pi^{2}}{12}
\frac{T^{2}}{\gamma_{0}^{2}}\right],
\label{oresults2b}
\\
\frac{\partial\chi^{\prime\prime}_{B_{2g}}}{\partial\Omega}
(T\ll T^{*})
= \frac{2N_{F}}{\pi\Delta_{0}}
b_{2}^{2}\left[1+\frac{\pi^{2}}{36}\frac{T^{2}}{\gamma_{0}^{2}}\right],
\label{oresults1a}
\\
\frac{\partial\chi^{\prime\prime}_{B_{1g}}}{\partial\Omega}
(T\ll T^{*})
= \frac{2N_{F}}{\pi\Delta_{0}}
b_{1}^{2}\frac{\gamma_{0}^{2}}{\Delta_{0}^{2}} \ln(4\Delta_{0}/\gamma_{0})\nonumber\\
\times\left[1-\frac{\pi^{2}}{12}\frac{T^{2}}{\gamma_{0}^{2}}\right],
\label{oresults1b}
\end{eqnarray}
where $n$ is the 2D electron density.
Eq. (\ref{oresults2a}) for the in-plane conductivity has been
derived several times \cite{Hirschfeld94,Graf95,Graf96,DurstLee},
and Eqs. (\ref{oresults1a}-\ref{oresults1b}) for the Raman slope
are identical to those found in Ref. \cite{WuCarbotte}. The result
for the out-of-plane conductivity for $T=0$ is also in agreement
with the result from Ref. \cite{KimCarbotte}, but the temperature
dependent variation has not been presented before. We note as in
Refs. \cite{WuCarbotte,Hirschfeld94,Graf95,Graf96,DurstLee,tpdapk}
that both the in-plane conductivity and the $B_{2g}$ Raman slope
are universal numbers for resonant scattering independent of the
strength of the scattering, while both the c-axis conductivity and
the $B_{1g}$ Raman slope depend on $\gamma_{0}$. The $\gamma_{0}$
dependence does not appear in the $c-$axis conductivity if the
$c-$axis hopping is taken as a constant independent of direction
around the FS \cite{Latyshev99,KimCarbotte}. The temperature
dependencies are {\it positive} for both the in-plane conductivity
and the $B_{2g}$ slope, but are {\it negative and identical} for the
out-of-plane conductivity and the $B_{1g}$ slope, giving a peak at
zero $T$ for the latter pair. We note that this result is in
agreement with the rise of the $c-$axis conductivity recently
observed in YBa$_{2}$Cu$_{3}$O$_{6.95}$ at low
temperatures\cite{Hosseini98}.

In the limit of higher temperatures $T_{c}\gg T \gg T^{*}$ where
the DOS does not have an impurity induced weight at
the Fermi level and matches the DOS from the clean limit, the self
consistency is not required for the self energy and Eq.
(\ref{superconducting}) can be rewritten as
\begin{eqnarray}
&&\left(
\begin{array}{c}
\sigma_{\alpha,\beta}^{\prime}(T^{*}\ll T \ll T_{c})\\
\partial \chi^{\prime\prime}_{\gamma,\gamma}(T^{*}\ll T \ll T_{c})/\partial \Omega\\
\end{array}
\right)
=\nonumber\\
&&-2N_{F}\int dx {\partial f(x)\over{\partial x}}
Im\left[{1\over{\Omega-i/\tau(x)}}\right]\nonumber\\
&&\times\Biggl\langle
\left(
\begin{array}{c}
v_{\alpha}({\bf k})v_{\beta}({\bf k})
\\
\gamma^{2}({\bf k})\\
\end{array}
\right)
Re\left[{x\over{\sqrt{x^{2}-\Delta({\bf k})^{2}}}}\right]
\Biggr\rangle,
\label{Drude}
\end{eqnarray}
with $1/\tau(x)=-2\Sigma^{\prime\prime}_{0}(x)$. This is a generalization of the
results obtained in Refs. \cite{XiangWheatley,XiangHardy,Hirschfeld94} to the case of
Raman and optical conductivity. We note that for $d-$ wave superconductors in the
resonant limit, the impurity scattering rate depends strongly on frequency
\begin{equation}
1/\tau(\omega\rightarrow 0)=\frac{\pi^{2}\Gamma\Delta_{0}}{2\omega}
\frac{1}{\ln^{2}(4\Delta_{0}/\omega)},
\label{tau}
\end{equation}
as shown in Ref. \cite{Hirschfeld94}, which yields
\begin{eqnarray}
\left(
\begin{array}{c}
\sigma_{\alpha,\beta}^{\prime}(T^{*}\ll T\ll T_{c})\\
\partial \chi^{\prime\prime}_{\gamma,\gamma}(T^{*}\ll T \ll T_{c})/\partial \Omega\\
\end{array}
\right)
=\nonumber\\
-\frac{4N_{F}}{\pi^{2}}\frac{T^{2}}{\Gamma\Delta_{0}}
\int dz z^{2}\frac{e^{z}}{(e^{z}+1)^{2}}\nonumber\\
\times \ln^{2}(4\Delta_{0}/zT)
H^{\sigma_{\alpha,\beta},\chi_{\gamma,\gamma}}(zT),
\label{hight}
\end{eqnarray}
with the functions
\begin{eqnarray}
&&
H^{\sigma_{\alpha,\beta}}(x)
=Re\biggl\langle
\frac{v_{\alpha}({\bf k})v_{\beta}({\bf k})}{\sqrt{x^{2}-\Delta({\bf k})^{2}}}
\biggr\rangle,
\\
&&H^{\chi_{\gamma,\gamma}}(x)
=Re\biggl\langle
\frac{\gamma^{2}({\bf k})}{\sqrt{x^{2}-\Delta({\bf k})^{2}}}
\biggr\rangle.
\end{eqnarray}
Performing the integrals gives for small $x$ gives
\begin{eqnarray}
H^{\chi_{\gamma,\gamma}}(x)=
\cases{\frac{x^{2}}{2\Delta_{0}^{3}},~~~B_{1g},\cr
\frac{1}{\Delta_{0}},~~~B_{2g},}\\
H^{\sigma_{\alpha,\beta}}(x)=
\cases{\frac{1}{2\Delta_{0}},~~~\sigma_{xx},\cr
\frac{x^{2}}{4\Delta_{0}^{3}},~~~\sigma_{zz}.}
\end{eqnarray}
The remaining integrals in Eq. (\ref{hight}) can be easily performed:
\begin{eqnarray}
\sigma_{\alpha,\beta}^{\prime}(T^{*}\ll T \ll T_{c})=\nonumber\\
\cases{\frac{2ne^{2}}{3m\Gamma}\left(\frac{T}{\Delta_{0}}\right)^{2}
\ln^{2}\left(\frac{4\Delta_{0}}{T}\right), ~~~\sigma_{xx}, \cr
\frac{14\pi^{2}ne^{2}v_{z}^{2}}{15m\Gamma v_{F}^{2}}
\left(\frac{T}{\Delta_{0}}\right)^{4}
\ln^{2}\left(\frac{4\Delta_{0}}{T}\right),~~~\sigma_{zz},}\\
\frac{\partial \chi^{\prime\prime}_{\gamma,\gamma}(T^{*}\ll T \ll T_{c})}
{\partial \Omega}=\nonumber\\
\cases{\frac{4b_{2}^{2}N_{F}}{3\Gamma}\left(\frac{T}{\Delta_{0}}\right)^{2}
\ln^{2}\left(\frac{4\Delta_{0}}{T}\right),~~~B_{2g},\cr
\frac{14\pi^{2}b_{1}^{2}N_{F}}{15\Gamma}
\left(\frac{T}{\Delta_{0}}\right)^{4}
\ln^{2}\left(\frac{4\Delta_{0}}{T}\right),~~~B_{1g}.}
\label{hightr}
\end{eqnarray}

\begin{figure}
\psfig{file=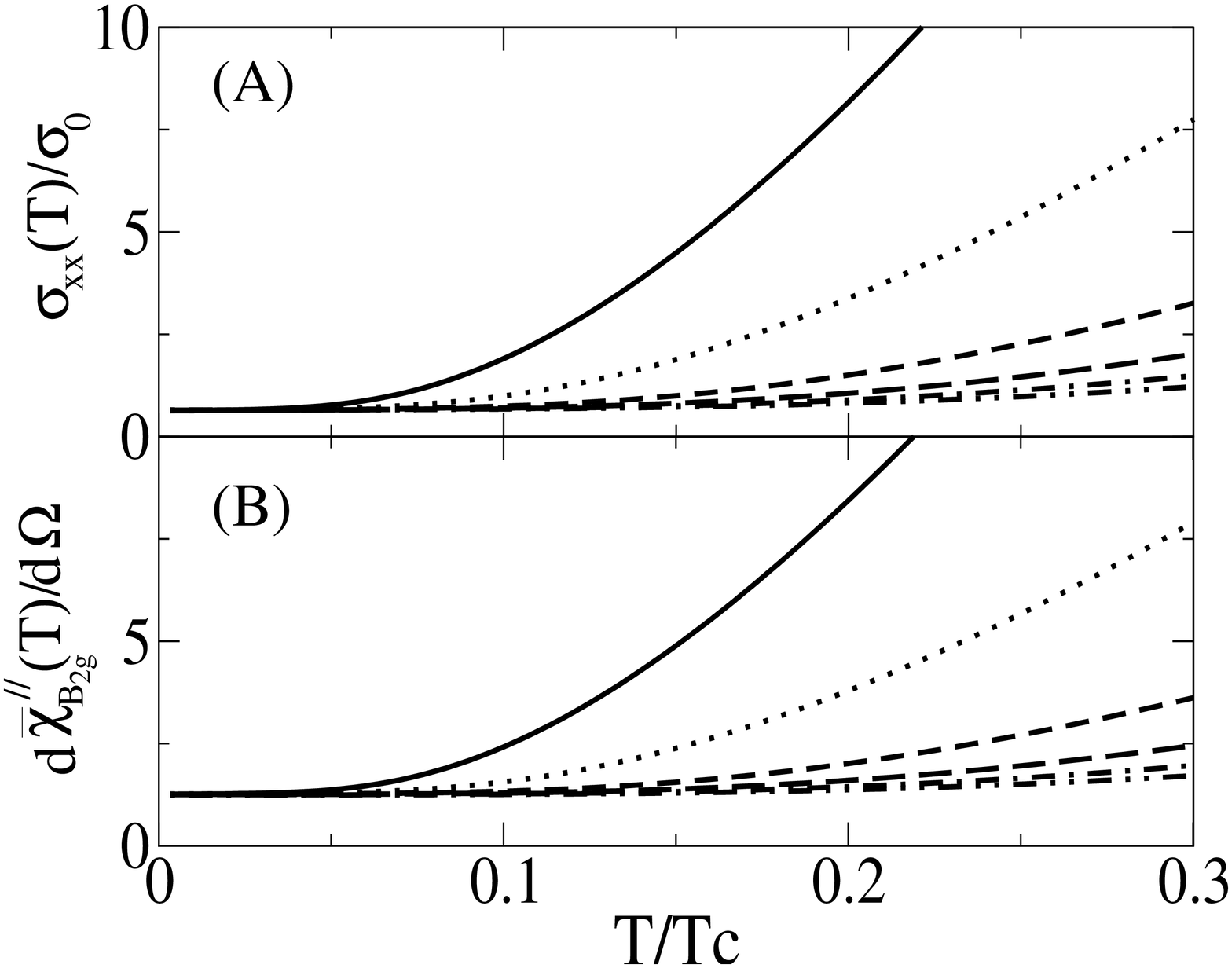,height=7.0cm,width=8.0cm,angle=0}
\caption[]{
Temperature dependence of the in-plane conductivity (Panel A) and
the $B_{2g}$ Raman slope (Panel B) for resonant scattering and different
impurity
scattering strengths $\Gamma/\Delta_{0}=0.004, 0.008, 0.016, 0.024$ and $0.04$
(solid, dotted, short-dashed, long-dashed, and dotted-dashed lines),
respectively, for $\Delta_{0}/T_{c}=4$. Here $\sigma_{0}=\pi N_{F}e^{2}v_{F}^{2}/\Delta_{0}$
and $\bar\chi^{\prime\prime}=\chi^{\prime\prime}/N_{F}b_{2}^{2}$ are the
dimensionless quantities shown.}
\label{fig4}
\end{figure}

The expression for the in-plane conductivity was derived in Ref. \cite{Hirschfeld94}
but to our knowledge the other terms are new. We note the results for $\sigma_{zz}$
and $\sigma_{xx}$ in this limit differ from those obtain in Ref. \cite{XiangHardy},
where a frequency independent scattering time was used
rather than that of Eq. (\ref{tau}). As
a consequence they concluded that $\sigma_{xx,zz}(T)\propto n_{xx,zz}(T)$ with
$n_{xx,zz}(T)$ the normal-fluid density which decreases uniformly with temperature
in contrast to experiments\cite{Hosseini98}. From that they concluded that the
scattering time must be anisotropic. We note that any frequency dependence of
the scattering time would qualitatively change this conclusion.

\begin{figure}
\psfig{file=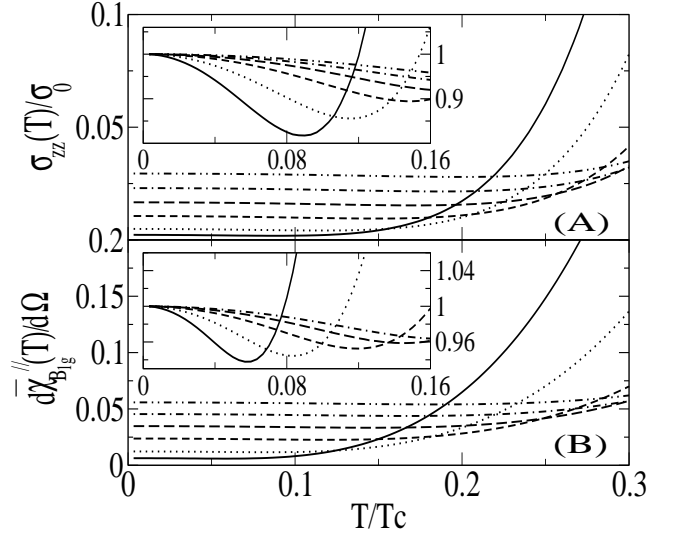,height=7.0cm,width=8.5cm,angle=0}
\caption[]{
Temperature dependence of the out-of-plane conductivity (Panel A) and
the $B_{1g}$ Raman slope (Panel B) for resonant scattering (c=0) and different
impurity
scattering strengths $\Gamma/\Delta_{0}=0.004, 0.008, 0.016, 0.024$ and $0.04$
(solid, dotted, short-dashed, long-dashed, and dotted-dashed lines),
respectively, for $\Delta_{0}/T_{c}=4$. Here $\sigma_{0}=\pi N_{F}e^{2}v_{z}^{2}/\Delta_{0}$
and $\bar\chi^{\prime\prime}=\chi^{\prime\prime}/N_{F}b_{1}^{2}$ are the
dimensionless quantities shown.
Insets: low temperature
rise of both $\sigma_{zz}$ and $\partial\chi^{\prime\prime}/\partial\Omega$
(normalized to their zero temperature values) with decreasing temperature.}
\label{fig5}
\end{figure}

The results of Eqs. (\ref{oresults2a}-24) and (32-\ref{hightr})
imply that the SS relations in the normal state Eq. (\ref{genSS})
hold in the superconducting state. The exponent of the low
temperature rise {\it as well as the sign of the correction} do
obey the general scaling relation, following simply from the
interplay of anisotropies of $\Delta({\bf k})$ and the respective
vertices.

Again these relations Eqs. (\ref{oresults2a}-24) and (32-\ref{hightr}) would be expected to
be violated for the same reasons discussed in the normal state in Section II. However
additional considerations should be mentioned here as well. It is well known that at low temperatures
$T\ll T^{*}$ the $ab$-plane conductivity in Y-123 varies as $T^{\alpha}$ with an exponent $\alpha\le 1$\cite{Bonn93} and
not $T^{2}$ predicted by Eq. (\ref{oresults2a}), and has been found generally to be non-universal in the zero $T$ 
limit\cite{Berlinsky}.  While vertex corrections can address non univeral numbers\cite{DurstLee} 
and scattering away from the unitary limit changes $\alpha$ from 1\cite{Hensen97,Berlinsky},
systematic agreement has not been reached at low temperatures. To address this discrepancies, recently Atkinson and
Hirschfeld\cite{Atkinson} have shown that a reduced $ab$-plane conductivity emerges at low temperatures
when real-space variations of the order parameter in the neighborhood of the impurities and impurity interference
effects are consider in a semiclassical Bogolubov - de Gennes framework. These effects are not captured in the self
consistent $T-$matrix approach and are thus beyond the scope of the present manuscript. It is not immediately clear
how the changes in $\sigma_{ab}(T)$ are manifest in other response functions considered in this manuscript and how
the derived scaling relations are affected. Our approach should be valid at not too low
temperatures where deviations of the conductivity from the unitary limit results are found.

The response for $T\ll T_{c}$ is calculated by numerically solving
Eq. \ref{superconducting} and the corresponding self-consistent
equations to determine the self energies. The results for
$\sigma_{xx}(T), \partial\chi^{\prime\prime}_{B_{2g}}(T)/\partial
\Omega$, and $\sigma_{zz}(T),
\partial\chi^{\prime\prime}_{B_{1g}}(T)/\partial \Omega$ are shown
in Figures \ref{fig4} and \ref{fig5}, respectively, for resonant
scattering and different values of the impurity scattering
strengths $\Gamma/\Delta_{0}$. Generally at higher temperatures
$T>T^{*}$ all quantities increase rapidly with temperature, rising
as $T^{2}$ and $T^{4}$ for $\sigma_{xx},
\partial\chi_{B_{2g}}/\partial \Omega$ and $\sigma_{zz},
\partial\chi_{B_{1g}}/\partial \Omega$, respectively. The rise of
the $c-$axis conductivity and the $B_{1g}$ Raman slope at low
temperatures shown in the insets of Fig. \ref{fig5} are generally
on the order of a few percent for the parameters shown. This
height rises for smaller values of $\Gamma$ but onsets at smaller
temperatures due to the concomitant reduction in $T^{*}$. In
particular, the rise and the onset of the $c-$axis conductivity
low temperature maximum for YBa$_{2}$Cu$_{3}$O$_{6.95}$
\cite{Hosseini98} cannot be adequately reproduced. There are as
yet no Raman measurements to compare to, and thus it would be
extremely useful to have data on a wide range of compounds and
doping levels as well as a systematic check of impurity doping
effects to test these results.

\subsection{Spin fluctuations}

The different rate of descent of the response functions below
$T_{c}$ has an interesting consequence on the conductivity peak
seen in $ab$-plane measurements and the lack of peak seen in
$c-$axis measurements. Ref. \cite{Hirschfeld94} included inelastic
scattering from spin fluctuations in RPA to reproduce the
$ab$-peak in the conductivity observed in Ref.
\cite{Bonn93,Hosseini99}. In Refs. \cite{Hirschfeld94,XiangHardy}
it was shown that Eq. (\ref{Drude}) for the conductivities for
$T_{c}\gg T\gg T^{*}$ may be reexpressed in terms of the normal qp
density which can be generalized as
\begin{equation}
\sigma_{\alpha,\beta}(T)=\frac{n_{qp}^{\alpha\beta}(T)e^{2}}{m}
\bar\tau,
\label{condnq}
\end{equation}
with
\begin{equation}
n_{qp}^{\alpha,\beta}(T)=\frac{1}{v_{F}^{2}}
\int d\omega\biggl\langle
Re \left[
\frac{v_{\alpha}({\bf k})v_{\beta}({\bf k})
\omega}{\sqrt{\omega^{2}-\Delta({\bf k})^{2}}}
\right]
\biggr\rangle
\left[-\frac{\partial f}{\partial\omega}\right]
\end{equation}
the projected normal quasiparticle density.
The average $\bar\tau$ is derived from the frequency-dependent
$\tau(\omega)$ and the superconducting DOS
$N(\omega)$ as
\begin{equation}
\bar \tau = \frac{\int d\omega N(\omega)(-\partial f/\partial \omega)\tau(\omega)}
{\int d\omega N(\omega)(-\partial f/\partial\omega)}.
\end{equation}
Similarly one can re-express the Raman slopes in the same fashion:
\begin{equation}
\partial\chi^{\prime\prime}_{\gamma,\gamma}(T)/\partial\Omega
=n_{qp}^{R,\gamma\gamma}(T)\bar\tau,
\label{Ramannq}
\end{equation}
with
\begin{equation}
n_{qp}^{R,\gamma\gamma}(T)=
\int d\omega\biggl\langle
Re \left[
\frac{\gamma^{2}({\bf k})\omega}{\sqrt{\omega^{2}-\Delta({\bf k})^{2}}}
\right]
\biggr\rangle
\left[-\frac{\partial f}{\partial\omega}\right]
\end{equation}
the Raman projected normal qp density. For a $d_{x^{2}-y^{2}}$
gap, from the results of Eqs. (28-31)
the projected qp densities at low $T$ vary as $T$ for
$n_{qp}^{xx}, n_{qp}^{R,B_{2g}}$ and $T^{3}$ for
$n_{qp}^{zz}, n_{qp}^{R,B_{1g}}$, respectively. If the scattering time
$\tau$ were independent of frequency, then $n_{qp}$ gives the full
temperature dependence and thus $\sigma_{xx},\partial\chi^{\prime\prime}_{B_{2g}}
/\partial\Omega$ would vary linearly with $T$ and
$\sigma_{zz},\partial\chi^{\prime\prime}_{B_{1g}}
/\partial\Omega$ would vary as $T^{3}$. Ref. \cite{XiangHardy} used
this result for $\sigma_{zz}$ and argued that $T^{3}$ accurately fit the data
for $T> 40 K$, but they could not explain the rise at low $T$. However, the
impurity scattering rate as well as the scattering due to inelastic collisions,
such as spin fluctuations, depend on momentum and strongly depend
on both temperature and
frequency. The latter is crucially needed in order to explain the peak in the $ab$-plane
conductivity.

\begin{figure}
\psfig{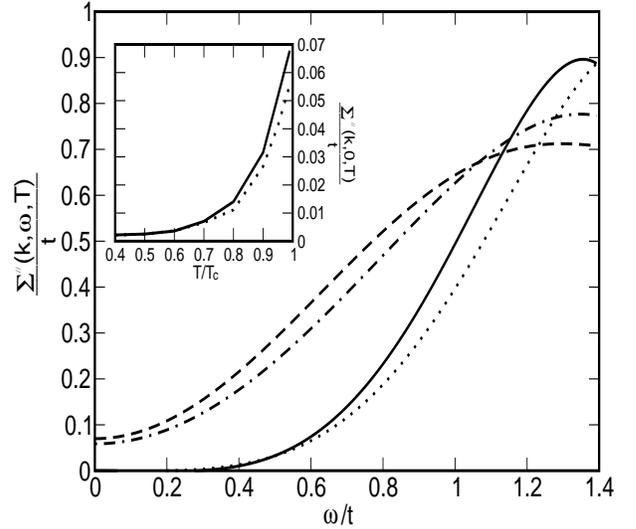}
\caption[]{Frequency dependence of the
imaginary part of the $\hat\tau_{0}$ self energy
$\Sigma^{\prime\prime}_{0}({\bf k},\omega,T)$ normalized to the
hopping overlap $t$ as a function of
frequency and temperature for different points in the BZ.
The solid line and dotted line are for $T=0.5T_{c}$ for gap maximum
${\bf k}=(\pi/a,0)$ and gap node
$(\pi/2a,\pi/2a)$ point, respectively, while the dashed and dot-dashed lines
correspond to the gap max and gap node points at $T_{c}$. The
inset shows the zero frequency part of
$\Sigma^{\prime\prime}_{0}$ as a function of temperature.}
\label{fig9}
\end{figure}

Refs. \cite{Hirschfeld94,Hirschfeld97,tpdapk2000} utilized calculations of the
inelastic scattering due to spin fluctuations in the
2D Hubbard model in the Random Phase approximation (RPA) for $U=2t$
to describe the dc and IR conductivity and the frequency dependent Raman
response. The lifetime calculated for $U=2t$ and $\Delta_{0}/T_{c}=3-4$\cite{Quinlan94}
was found to give reasonable agreement with the transport lifetime
determined from conductivity measurements in Y-123\cite{Bonn93} and gave
reasonable agreement with the $ab-$plane conductivity peak\cite{Hirschfeld94},
$ab$-plane IR conductivity response\cite{Hirschfeld97}, and simultaneously
the $ab-$plane IR and the
$B_{1g}$ and $B_{2g}$ Raman response in Bi-2212\cite{tpdapk2000}. We therefore
use this approach to calculate the temperature dependence of the response functions
for all temperatures below $T_{c}$.

In RPA, the self energy $\Sigma_{0}$ is given from the effective potential
$V$ as:
\begin{equation}
V({\bf q},i\Omega)={3\over{2}}
{{\bar U}^{2}\chi_{0}({\bf q},i\Omega)\over{1-\bar U
\chi_{0}({\bf q},i\Omega)}},
\label{eq5}
\end{equation}
where $\bar U$ is a phenomenological parameter [we choose $\bar U=2t$].
$\chi_{0}({\bf q},i\Omega)$ is the non-interacting spin susceptibility,
\begin{widetext}
\begin{equation}
\chi_{0}({\bf q},i\Omega)=\sum_{\bf k}
\biggl\{{a^{+}_{\bf k,k+q}\over{2N}}
{f(E_{\bf k+q})-f(E_{\bf k})\over{i\Omega-(E_{\bf k+q}-E_{\bf k})}}
+ {a^{-}_{\bf k,k+q}\over{4N}}
\left[{1-f(E_{\bf k+q})-f(E_{\bf k})\over{i\Omega+E_{\bf k+q}+E_{\bf k}}}
-{1-f(E_{\bf k+q})-f(E_{\bf k})\over{i\Omega-E_{\bf k+q}-E_{\bf k}}}\right]
\biggr\}.
\label{eq6}
\end{equation}
Here
$E_{\bf k}^{2}=\epsilon_{\bf k}^{2}+\Delta_{\bf k}^{2}$
and the coherence factors are
$a^{\pm}_{\bf k,k+q}=1\pm {\epsilon_{\bf k+q}\epsilon_{\bf k}+
\Delta_{\bf k}\Delta_{\bf k+q}\over{E_{\bf k+q}E_{\bf k}}}$.
This yields a self energy
\begin{eqnarray}
&&\hat\Sigma({\bf k},i\omega)=-\int{dx\over{\pi N}}
\sum_{\bf q} {V^{\prime\prime}({\bf q},x)\over{2E_{\bf k-q}}}
\\
&&\times
\Biggl[{E_{\bf k-q}\hat\tau_{0}+\epsilon_{\bf k-q}\hat\tau_{3}+
\Delta_{\bf k-q}\hat\tau_{1}\over{E_{\bf k-q}+x-i\omega}}
[n(x)+f(-E_{\bf k-q})]
-{-E_{\bf k-q}\hat\tau_{0}+\epsilon_{\bf k-q}\hat\tau_{3}+
\Delta_{\bf k-q}\hat\tau_{1}\over{-E_{\bf k-q}+x-i\omega}}
[n(x)+f(E_{\bf k-q})]\Biggr],
\label{eq7new}
\nonumber
\end{eqnarray}
with $n$ the Bose factor.
\end{widetext}

The imaginary part of the $\hat\tau_{0}$ self energy
$\Sigma^{\prime\prime}_{0}({\bf k},\omega,T)$ normalized to the
hopping overlap $t$ as a function of
frequency and temperature for different points in the BZ
is shown in Fig. (\ref{fig9}). Here we have used the band
structure $\epsilon_{\bf k}$ as in (A4) in the appendix with
$t^{\prime}/t=0.45$ and a filling $n=0.88$, $U=2t$, and
a $d_{x^{2}-y^{2}}$
energy gap $\Delta_{\bf k}=\Delta_{0}[\cos(k_{x}a)-\cos(k_{y}a)]/2$
with $\Delta_{0}/t=0.4=4T_{c}/t$.
The solid line and dotted line
shows the frequency dependence of $\Sigma^{\prime\prime}$ at a
temperature $T=0.5T_{c}$ for gap maximum
${\bf k}=(\pi/a,0)$ and gap node
$(\pi/2a,\pi/2a)$, respectively, while the dashed and dot-dashed lines
correspond to the gap max and gap node points at $T_{c}$. The
differences for the gap maximum and gap node points are not too strong
and can be adequately fit with a threshold behavior
$\sim (\omega-3\Delta({\bf k}))^{3}$ plus a temperature dependent part which
also depends on momentum. The inset shows the zero frequency part of
$\Sigma^{\prime\prime}_{0}$ as a function of temperature. Except for
low temperatures where the nodal properties of the interaction govern
the behavior, the momentum dependence of the self energy is weak and
can be adequately modelled by a temperature dependent $\sim T^{3}$
term plus a frequency dependent part $\sim \omega^{3}$.

\begin{figure}
\psfig{file=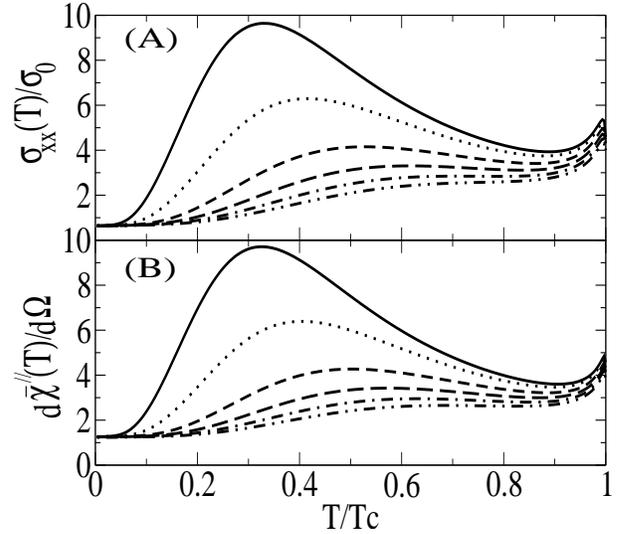,height=7.0cm,width=8.0cm,angle=0}
\caption[]{
Temperature dependence of the in-plane conductivity (Panel A) and
the $B_{2g}$ Raman slope (Panel B) including inelastic spin fluctuations and
resonant impurity scattering for different
impurity
scattering strengths $\Gamma/\Delta_{0}=0.004, 0.008, 0.016, 0.024$ and $0.04$
(solid, dotted, short-dashed, long-dashed, and dotted-dashed lines),
respectively, for $\Delta_{0}/T_{c}=4$. Here $\sigma_{0}=\pi N_{F}e^{2}v_{F}^{2}/\Delta_{0}$
and $\bar\chi^{\prime\prime}=\chi^{\prime\prime}/N_{F}b_{2}^{2}$ are the
dimensionless quantities shown.}
\label{fig7}
\end{figure}

\begin{figure}
\psfig{file=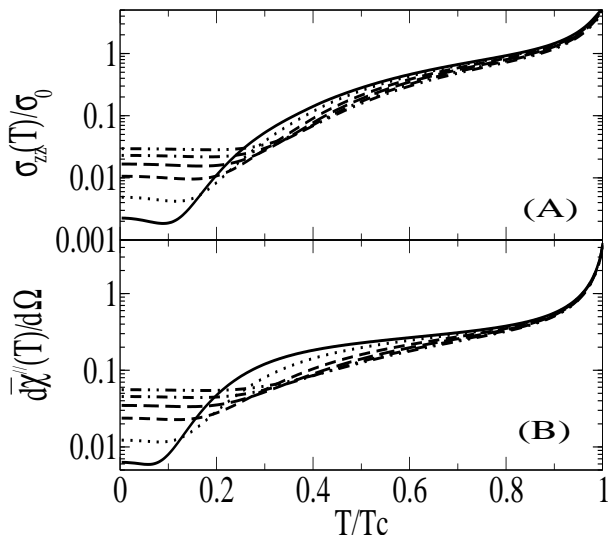,height=7.0cm,width=8.0cm,angle=0}
\caption[]{
Temperature dependence of the out-of-plane conductivity $\sigma_{zz}$ (Panel A) and
the $B_{1g}$ Raman slope (Panel B) including inelastic spin fluctuations and
resonant impurity scattering for different impurity
scattering strengths $\Gamma/\Delta_{0}=0.004, 0.008, 0.016, 0.024$ and $0.04$
(solid, dotted, short-dashed, long-dashed, and dotted-dashed lines),
respectively, for $\Delta_{0}/T_{c}=4$. Here $\sigma_{0}=\pi N_{F}e^{2}v_{F}^{2}/\Delta_{0}$
and $\bar\chi^{\prime\prime}=\chi^{\prime\prime}/N_{F}b_{1}^{2}$ are the
dimensionless quantities shown.}
\label{fig8}
\end{figure}

In an effort to address the temperature dependence of these
quantities, we employ a simple parameterized fit to the numerical
results for
$1/\tau_{\bf k}(\omega,T)=-2\Sigma_{0}^{\prime\prime}({\bf k},\omega, T)$
determined from Eq. (\ref{eq7new}) and Fig. (\ref{fig8})
and add that to the elastic contribution
calculated in the last section.
Assuming Matthiessen's law to hold
in this case neglects vertex corrections and the joint influence
of disorder on the spin fluctuations and vice-versa, but for weak
disorder should be sufficient to capture the qualitative behavior
of various quantities derived on the FS.

The results for the four response functions derived from Eq.
(\ref{superconducting}) are shown in Figures
\ref{fig7} and \ref{fig8}. Both the in-plane conductivity (Fig.
\ref{fig7}A) and the $B_{2g}$ Raman slope (Fig. \ref{fig7}B)
possess a peak near $T\sim 0.3 T_{c}$ for
$\Gamma/\Delta_{0}=0.004$ which decreases in height and moves to
higher temperatures for increasing impurity scattering. It is
important to emphasize that this peak is not related to coherence
effects and is a simple balance of fall-off of the inelastic
scattering rate $\sim T^{3}$ and the rise of the impurity
scattering rate $\sim 1/T$ at low temperatures. The sum of these
scattering rates is multiplied by the normal qp density $\propto
T$ at low temperatures, as shown in Eqs. (\ref{hight}) and
(\ref{hightr}), and therefore $\sigma_{xx}$ and
$\partial\chi^{\prime\prime}_{B_{2g}}/\partial \Omega$ vary as
$1/T^{2}$ for $T^{*}<T < T_{c}$ and approach a universal constant
in the zero temperature limit.

However, no corresponding
peak is found for both the out-of-plane conductivity (Fig. \ref{fig8}A)
and $B_{1g}$ Raman slope (Fig. \ref{fig8}B), in agreement with experimental observations.
The curves simply show a rapid fall off
of both quantities for $T<T_{c}$ and a small rise of both quantities which onsets at
$T^{*}$ and reaches a zero temperature maximum as shown in Fig. \ref{fig5}. The main difference is due to
the behavior of zone-axis projected qp density, which varies as $T^{3}$ at low
temperatures, with a factor of $T$ coming from the nodes and the additional $T^{2}$ coming
from the matrix elements. This compensates the $1/T^{3}$ rise of the qp inelastic lifetime,
and both
$\sigma_{zz}$ and
$\partial\chi^{\prime\prime}_{B_{1g}}/\partial \Omega$ vary as $T^{\mu}$ for
$T\gg T^{*}$ with $\mu$ dependent on the strength of the impurity scattering,
and rise for $T\ll T^{*}$, as shown in Fig. (\ref{fig5}). For example, for the parameters
chosen in Fig. (\ref{fig8}), the exponent $\mu$ for $0.3T_{c}<T<0.9T_{c}$ for
the $c-$axis conductivity $\sigma_{zz}(T)$ varies from 2.7 to 3.4 for increasing impurity
scattering. If the frequency dependence of the scattering rate were neglected, then a
universal exponent $\mu=3$ would emerge\cite{XiangHardy}. Therefore it would be highly useful
if further systematic checks were performed and Raman data were available to compare to
the conductivity and the theoretical predictions.

\section{Summary and conclusions}

In summary, based on symmetry arguments we have demonstrated how
the relaxational behavior of the qp in the cuprates should
manifest itself in the various experiments and how the results are
expected to be interrelated. Therefore, a single framework may
relate the optically-derived qp scattering rates to transport
measurements to infer charge dynamics on different regions of the
Brillouin zone. Using forms for the interlayer hopping and qp self
energy consistent with empirical evidence, a variant of the SS
relation was shown to relate the zone-diagonal and zone-axis
transport properties measured by DC conductivity and the slope of
the Raman response in the normal state, in agreement with
experimental observations in Bi-2212 and Y-123, but not La-214.
Violations of the derived scaling relations were discussed most
pointedly in connection with the role of vertex corrections.

The ``scattering ratios'' show power-law behavior for the
Raman response which can be reasonably accounted for in several
models near optimal doping. However no single model can adequately
describe the data over the entire doping range, indicating that
additional physics related to strong correlations is required\cite{QCP}.
A presence of a pseudogap is discussed in simple symmetry terms, revealing that the $B_{1g}$
Raman scattering and c-axis conductivity are most affected in agreement with experiments.
This is a consequence of a loss of qp coherence near the BZ axes.

The data on La-214 over a wide range of doping is inconsistent with the simple models for qp
scattering discussed herein. A connection can be made between the in-plane
conductivity and Raman response in light of stripe orientation. However more work is clearly needed to
address this point.

We note that a quantitative connection between the magnitude and temperature dependence of the
qp self energy derived from ARPES, the in-plane and out-of-plane conductivity, and the Raman response
can only be undertaken with an understanding of the role of vertex corrections. 

In the superconducting state, a similar SS relation is found which arises
from the momentum dependence of the energy gap and conductivity and Raman
matrix elements. In particular, we found that the a zero temperature peak is predicted
to arise in $\sigma_{zz}$ and $\partial\chi^{\prime\prime}_{B_{1g}}/\partial\Omega$
without the presence of a maximum near $0.3T_{c}$ found for $\sigma_{xx}$
and $\partial\chi^{\prime\prime}_{B_{1g}}/\partial\Omega$. The results are in
rough, qualitative agreement with the available data for $\sigma_{zz}$ but
the strength of the elastic scattering cannot simultaneously account for in-plane
and out-of-plane conductivities. However, the simple model presented does not
account for anisotropies in impurity scattering, known to arise for point-like
scatterers in correlated materials, or impurity interference effects. In particular, it would useful to determine
whether the approach followed by Atkinson and Hirschfeld\cite{Atkinson} would remedy the agreement to the $c-$axis
conductivity rise at low temperatures. Unfortunately,
Raman data in the superconducting state to further test the theory is lacking.
In particular, it would
be extremely useful to determine if the deviations from the derived SS relation observed
in the normal state of La-214 carry over into the superconducting state.

The agreement of the derived SS relations in both the superconducting and
normal states with the available data on Bi-2212 and Y-123 indicate that
the in-plane momentum is at least partially conserved
in c-axis transport over the entire doping range studied.
This shows that in principle a comparison of Raman and transport
could eventually contribute to the solution of the c-axis transport problem.

\acknowledgements
The author would like to thank R. Hackl, F. Venturini, G. Blumberg, A. Kampf and X.-G. Wen
for numerous discussions, and gratefully acknowledges support from NSERC, PREA and from the
Alexander von Humboldt Foundation.

\appendix
\section{}
We start by considering a four band model for the $CuO_{2}$ plane
with $Cu_{3d}-O_{p_{x,y}}$ hopping amplitude $t_{pd}$, $Cu_{4s}-O_{p_{x,y}}$
hopping amplitude $t_{ps}$, $O_{p_{x}}-O_{p_{y}}$ hopping amplitude $t_{pp}$, and
c-axis Cu$_{4s}$-Cu$_{4s}$ amplitude $t_{s}$, respectively:

\begin{eqnarray}
H&=&\epsilon_{d}\sum_{{\bf n},\sigma}d^{\dagger}_{{\bf n},\sigma}
d_{{\bf n},\sigma}+\epsilon_{s}\sum_{{\bf n},\sigma}s^{\dagger}_{{\bf n},\sigma}
s_{{\bf n},\sigma}\label{eq1} \nonumber\\
&-&t_{pd}\sum_{{\bf n},\delta,\sigma}P_{\delta}
(d^{\dagger}_{{\bf n},\sigma}a_{{\bf n},\delta,\sigma}+h.c.)\nonumber\\
&-&t_{pp}\sum_{{\bf n},\delta,\delta^{\prime}\sigma}
P^{\prime}_{\delta,\delta^{\prime}}
a^{\dagger}_{{\bf n},\delta,\sigma}a_{{\bf n},\delta^{\prime},\sigma}\nonumber\\
&-&t_{ps}\sum_{{\bf n},\delta,\sigma}P_{\delta}^{\prime\prime}
(s^{\dagger}_{{\bf n},\sigma}a_{{\bf n},\delta,\sigma}+h.c.)
\nonumber\\
&-&t_{s}\sum_{\langle{\bf n,m}\rangle,\sigma}
s^{\dagger}_{{\bf n},\sigma}s_{{\bf m},\sigma},
\label{Hamiltonian}
\end{eqnarray}
where
$\epsilon_{s,d}=E_{s,d}-E_{p}$ represents the charge transfer energy from the oxygen $p-$
to Cu$_{4s,3d}$ orbitals, respectively.
Here $s^{\dagger}_{\bf n,\sigma},d^{\dagger}_{\bf n,\sigma}$
creates an $4s,3d_{x^{2}-y^{2}}$ electron,
respectively, with spin $\sigma$ at a copper lattice site ${\bf n}$,
while $a_{{\bf n,\delta}},\sigma$ annihilates an electron at one of the
neighboring oxygen sites ${\bf n+\delta}/2$ determined by the unit vector
${\bf \delta}$ assuming the four values, $(\pm 1,0)$ and $(0,\pm 1)$.
The overlap factors $P$ have the following properties: $P_{(1,0)}=
P^{\prime\prime}_{(1,0)}=1,
P_{(0,1)}=P^{\prime\prime}_{(0,1)}=-1,
P^{\prime}_{\bf x,y}=P^{\prime}_{\bf -x,-y}=1,
P^{\prime}_{\bf -x,y}=P^{\prime}_{\bf x,-y}=-1,$ respectively. Lastly the
bracket $\langle \cdots \rangle$ notes a sum over the nearest neighbor
Cu$_{4s}$ sites in the $c-$direction. Thus $c-$axis hopping is mediated by
the Cu$_{4s}$ orbitals hybridizing with the bonding and anti-bonding
$pd$ bands consistent with LDA\cite{oka}.

After Fourier transforming, the Hamiltonian is
$H=\sum_{{\bf k},\sigma}H_{{\bf k},\sigma}$
with
\begin{eqnarray}
H_{{\bf k},\sigma}=\epsilon_{d}d^{\dagger}_{{\bf k},\sigma}d_{{\bf k},\sigma}
+\epsilon_{s}({\bf k})s^{\dagger}_{{\bf k},\sigma}s_{{\bf k},\sigma}\nonumber\\
-\{2it_{pd}d^{\dagger}_{{\bf k},\sigma}[a_{x,{\bf k},\sigma}s_{x}({\bf k})-
a_{y,{\bf k},\sigma}s_{y}({\bf k})]+ h.c.\}\nonumber\\
-\{2it_{ps}s^{\dagger}_{{\bf k},\sigma}[a_{x,{\bf k},\sigma}s_{x}({\bf k})+
a_{y,{\bf k},\sigma}s_{y}({\bf k})]+ h.c.\}\nonumber\\
-4t_{pp}s_{x}({\bf k})s_{y}({\bf k})
[a^{\dagger}_{x,{\bf k},\sigma}a_{y,{\bf k},\sigma}+ h.c],
\label{Hamiltonian2}
\end{eqnarray}
with $s_{\alpha}({\bf k})=\sin(ak_{\alpha}/2)$ and
$\epsilon_{s}({\bf k})=\epsilon_{s}-2t_{ss}\cos(k_{z}c)$.
Eq. \ref{Hamiltonian2} can be
diagonalized by defining
``canonical fermions''\cite{shastry}:
\begin{eqnarray}
\alpha_{\bf k,\sigma}&=&i{s_{x}({\bf k})a_{x,{\bf k},\sigma}-
s_{y}({\bf k})a_{y,{\bf k},\sigma}\over{\mu({\bf k})}}\nonumber\\
\beta_{\bf k,\sigma}&=&-i{s_{y}({\bf k})a_{x,{\bf k},\sigma}+
s_{x}({\bf k})a_{y,{\bf k},\sigma}\over{\mu({\bf k})}},
\label{canonical1}
\end{eqnarray}
where $\mu({\bf k})^{2}=s_{x}^{2}({\bf k})+s_{y}^{2}({\bf k})$. This
gives anti-bonding, bonding bands hybridized with the Cu
orbitals. This four-band model can be reduced to an effective one-band model by
eliminating the $\beta$ band and the
two bands with high energies $\sim \epsilon_{s,d}$. This is achieved by
defining two other sets of canonical fermions and expanding in powers
of $t_{pd,pd,ss}/\epsilon_{s,d}$\cite{jefferson}. The single-band
dispersion is approximately given by
\begin{eqnarray}
\epsilon({\bf k})&=&-2t[\cos(k_{x}a)+\cos(k_{y}a)]
+4t^{\prime}\cos(k_{x}a)\cos(k_{y}a)\nonumber\\
&&-2t^{\prime\prime}
\cos(2k_{x}a)\cos(2k_{y}a)\nonumber\\
&&-t_{\perp}\cos(k_{z}c)[\cos(k_{x}a)-\cos(k_{y}a)]^{2}-\mu,
\label{bandstructure}
\end{eqnarray}
with the identification to lowest order of
$t=t_{pp}-t_{pd}^{2}/\epsilon_{d},
t^{\prime}=-t_{pp}/2+t_{ps}^{2}/8\epsilon_{s},
t^{\prime\prime}=t_{ps}^{2}/16\epsilon_{s},$ and
$t_{\perp}=t_{ss}t_{ps}^{2}/\epsilon_{s}^{2}$. This form for
the interplane hopping can also be derived in the framework of the
Hubbard model by projecting out the high-lying Cu $4s$ orbitals and
the high-lying $d-p$ spin triplets by solving the correlation problem
within the unit cell and treating the intercell hopping as a degeneracy
lifting perturbation\cite{XiangWheatley,shastry}.


\addcontentsline{toc}{section}{Bibliography}
\begin{thebibliography}{99}

\bibitem{ARPES}
A. Damascelli, Z. Hussain, and Z.-X. Shen
Rev. Mod. Phys. {\bf 75}, 473-541 (2003).

\bibitem{CooperGray}
S. L. Cooper and K. E.
Gray, in {\it Physical Properties of High-Temperature Superconductors IV},
edited by D. M. Ginsberg (World Scientific, Singapore, 1994).

\bibitem{caxis}
S. V. Dordevic, E. J. Singley, D. N. Basov, S. Komiya, Y. Ando, E. Bucher, C. C. Homes and M. Strongin,
Phys. Rev. B {\bf 65}, 134511 (2002);
S. Tajima, J. Sch\"utzmann, S. Miyamoto, I. Terasaki, Y. Sato, and R. Hauff,
Phys. Rev. B {\bf 55}, 6051 (1997).

\bibitem{Watanabe97}
T. Watanabe, T. Fujii, and A. Matsuda, Phys. Rev. Lett. {\bf 79}, 2113 (1997).

\bibitem{Kendziora93}
C. Kendziora, M. C. Martin, J. Haartge, L. Mihaly, and L. Forr\'o, Phys. Rev.
B {\bf 48}, 3531 (1993).

\bibitem{Forro93}
L. Forr\'o, Phys. Lett. A {\bf 179}, 140 (1993).

\bibitem{Spectral}
D. N. Basov, S. I. Woods, A. S. Katz, E. J. Singley, R. C. Dynes,
M. Xu, D. G. Hinks, C. C. Homes, and M. Strongin, Science {\bf 283}, 49 (1999);
D. N. Basov, C. C. Homes, E. J. Singley, M. Strongin, T. Timusk, G. Blumberg,
and D. van der Marel, Phys. Rev. B {\bf 63}, 134514 (2001);
H. J. A. Molegraaf, C. Presura, D. van der Marel, P. H. Kes, and M. Li, Science {\bf 295},
2239 (2002).

\bibitem{Tallon}
J. L. Tallon and J. W. Loram, Physica C {\bf 349}, 53 (2001).

\bibitem{AJL}
P. W. Anderson, {\it The Theory of Superconductivity in the High T$_{c}$
Cuprates} (Princeton University Press, Princeton, 1997); A. J. Leggett,
Phys. Rev. Lett. {\bf 83}, 392 (1999);
M. Turlakov and A. J. Leggett, Phys. Rev. B {\bf 63}, 64518 (2001).

\bibitem{IoffeMillis}
L. B. Ioffe and A. J. Millis, Science {\bf 285}, 1241 (1999); Phys. Rev. B {\bf 61}, 9077 (2000);
N. Shah and A. J. Millis, Phys. Rev. B {\bf 64}, 174506 (2001); {\bf 65}, 024506 (2001).

\bibitem{KimCarbotte}
W. Kim and J. P. Carbotte, Phys. Rev. B {\bf 63}, 054526 (2001);
E. Schachinger and J. P. Carbotte, Phys. Rev. B {\bf 64}, 94501 (2001).

\bibitem{Cornaglia01}
P. S. Cornaglia, K. Hallberg, and C. A. Balseiro, Phys. Rev. B {\bf 63}, 060504 (2001).

\bibitem{XiangWheatley}
T. Xiang and J. M. Wheatley, Phys. Rev. Lett. {\bf 77}, 4632 (1996).

\bibitem{XiangHardy}
T. Xiang and W. Hardy, Phys. Rev. B {\bf 63}, 024506 (2000).

\bibitem{vanderMarel}
D. van der Marel, Phys. Rev. B {\bf 60}, 765 (1999).

\bibitem{cold}
L. B. Ioffe and A. J. Millis, Phys. Rev. B {\bf 58}, 11631 (1998).

\bibitem{hot}
R. Hlubina and T. M. Rice, Phys. Rev. B {\bf 51}, 9253 (1995); D. Pines and
B. Stojkovi\'c {\bf 55}, 8576 (1997); {\bf 56}, 11931 (1997); R. Hlubina,
Phys. Rev. B {\bf 58}, 8240 (1998).

\bibitem{Coleman}
P. Coleman, A. J. Schofield, and A. M. Tsvelik, Phys. Rev. Lett. {\bf 76},
1324 (1996); A. T. Zheleznyak, V. M. Yakovenko, H. D. Drew, and I. I. Mazin, Phys.
Rev. B {\bf 57}, 3089 (1998);
{\bf 59}, 207 (1999); K. G. Sandeman and A. J. Schofield, Phys. Rev.
B {\bf 63}, 094510 (2001).

\bibitem{Varma}
E. Abrahams and C. M. Varma, Phys. Rev. Lett. {\bf 86}, 4652 (2001);
{\bf 88}, 139903 (2002).

\bibitem{abrik}
A. A. Abrikosov, Physica C {\bf 258}, 53 (1996).

\bibitem{atkin}
W. A. Atkinson and J. P. Carbotte, Phys. Rev. B {\bf 55}, 12748 (1997).

\bibitem{oka}
O. K. Andersen, O. Jepsen, A. I. Liechtenstein, and I. I. Mazin, Phys. Rev. B {\bf 49}, 4145 (1994);
J. Phys. Chem. Solids {\bf 56}, 1573 (1995).

\bibitem{Hackl}
R. Hackl, M. Opel, P. F. M\"uller, G. Krug, B. Stadlober, R. Nemetschek, H. Berger and
L. Forr\'o, J. Low Temp. Phys. {\bf 105}, 733 (1996).

\bibitem{Blumberg}
G. Blumberg and M.V. Klein, Journ. of Low Temp. Phys. {\bf 117}, 1001 (1999).

\bibitem{opel2000}
M. Opel, R. Nemetschek, C. Hoffmann, R. Philipp, P. F. M\"uller, R. Hackl, I. T\"utt\"o,
A. Erb, B. Revaz, E. Walker, H. Berger and L. Forr\'o,
Phys. Rev. B {\bf 61}, 9752 (2000).

\bibitem{QCP}
F. Venturini, M. Opel, T. P. Devereaux, J. K. Freericks, I. T\"utt\"o, B. Revaz, E. Walker, H. Berger,
L. Forr\'o, and R. Hackl, Phys. Rev. Lett.
{\bf 89}, 107003 (2002).

\bibitem{Homes}
C. C. Homes, T. Timusk, D. A. Bonn, R. Liang, W. N. Hardy, Physica C {\bf 254}, 265 (1995);
D. N. Basov, T. Timusk, B. Dabrowski, and J. D. Jorgensen, Phys. Rev. B {\bf 50}, 3511 (1994).

\bibitem{timusk}
see e.g., T. Timusk and B. Statt, Rep. Prog. Phys. {\bf 62}, 61 (1999).

\bibitem{Puchkov}
A. V. Puchkov, D. N. Basov, and T. Timusk, J. Phys.: Condens.Matter {\bf 8}, 10049 (1996);
N. L. Wang, P. Zheng, T. Feng, G. D. Gu, C. C. Homes, J. M. Tranquada, B. D. Gaulin, and T. Timusk,
Phys. Rev. B {\bf 67}, 134526 (2003).

\bibitem{Quilty}
J. W. Quilty, S. Tajima, S. Adachi, and A. Yamanaka
Phys. Rev. B {\bf 63}, 100508 (2001).

\bibitem{Nemet}
R. Nemetschek, M. Opel, C. Hoffmann, P. F. M\"uller, R. Hackl, H. Berger, L. Forr\'o, A.
Erb, and E. Walker, Phys. Rev. Lett. {\bf 78}, 4837 (1997).

\bibitem{Ramanold}
T. P. Devereaux, D. Einzel, B. Stadlober, R. Hackl, D. H. Leach, and J. J. Neumeier
Phys. Rev. Lett. {\bf 72}, 396-399 (1994).

\bibitem{tpdapk}
T. P. Devereaux and A. P. Kampf, Int. Journ. Mod. Phys. B {\bf 11},
2093 (1997).

\bibitem{B1gpair}
C. Kendziora and A. Rosenberg, Phys. Rev. B {\bf 52}, 9867-9870 (1995);
M. Kang, G. Blumberg, M. V. Klein, and N. N. Kolesnikov,
Phys. Rev. Lett. {\bf 77}, 4434 (1996);
X. K. Chen, J. G. Naeini, K. C. Hewitt, J. C. Irwin, R. Liang, and W. N. Hardy
Phys. Rev. B {\bf 56}, 513 (1997);
A. Sacuto, J. Cayssol, P. Monod, and D. Colson
Phys. Rev. B {\bf 61}, 7122 (2000);
M. F. Limonov, S. Tajima, and A. Yamanaka
Phys. Rev. B {\bf 62}, 11859 (2000);
S. Sugai and T. Hosokawa, Phys. Rev. Lett. {\bf 85}, 1112 (2000);
K. C. Hewitt and J. C. Irwin, Phys. Rev. B {\bf 66}, 054516 (2002).

\bibitem{Genoud}
J.-Y. Genoud, H. J. Trodahl, and A. E. Pantoja, Sol. State Commun.
{\bf113}, 285 (1999).

\bibitem{BlumbergScience}
G. Blumberg, M. Kang, M. V. Klein, K. Kadowaki, and C. Kendziora, Science {\bf 278}, 1427 (1997).

\bibitem{Hewitt}
K. C. Hewitt, N. L. Wang, J. C. Irwin, D. M. Pooke, A. E. Pantoja and H. J. Trodahl,
Phys. Rev. B {\bf 60}, 9943 (1999).

\bibitem{footnote}
A weak peak located near 600 cm$^{-1}$ in Bi-2212 appearing above $T_{c}$ has been a focus of attention
and mild controversy. Blumberg {\it et al.} interpreted the feature as a superconductivity related peak
lending creedence to a pre-formed pair scenario of the pseduogap\cite{BlumbergScience}. On the other hand,
Hewitt {\it et al.} investigated isotopical doped systems and found that the peak was
absent in oxygenated systems underdoped by Y substitutions\cite{Hewitt}, suggesting that this peak is of
phononic origin and related to oxygen disorder in the deoxygenated BiO layers.

\bibitem{Cooper}
P. Nyhus, S. L. Cooper, and Z. Fisk
Phys. Rev. B {\bf 51}, 15626 (1995); P. Nyhus, S. L. Cooper, Z. Fisk, and J. Sarrao
Phys. Rev. B {\bf 52}, 14308 (1995); {\it ibid.} {\bf 55}, 12488 (1997).

\bibitem{Hosseini99}
A. Hosseini, R. Harris, S. Kamal, P. Dosanjh, J. Preston, R. Liang,
W. N. Hardy, and D. A. Bonn, Phys. Rev. B {\bf 60}, 1349 (1999).

\bibitem{Bonn93}
D. A. Bonn, R. Liang, T. M. Riseman, D. J. Baar, D. C. Morgan, K. Zhang, P. Dosanjh, T. L. Duty,
A. MacFarlane, G. D. Morris, J. H. Brewer, W. N. Hardy, C. Kallin and A. J. Berlinsky,
Phys. Rev. B {\bf 47}, 11314 (1993).

\bibitem{Hirschfeld97}
P. J. Hirschfeld, S. M. Quinlan, and D. J. Scalapino, Phys. Rev. B
{\bf 55}, 12742 (1997).

\bibitem{Hirschfeld94}
P. J. Hirschfeld, W. O. Puttika, and D. J. Scalapino, Phys. Rev. Lett.
{\bf 71}, 3705 (1993); Phys. Rev. B {\bf 50}, 10254 (1994).

\bibitem{Hensen97}
S. Hensen, G. M\"uller, C. T. Rieck and K. Scharnberg, Phys. Rev. B {\bf 56},
6237 (1997).

\bibitem{WalkerDuffy}
M. B. Walker and M. F. Smith, Phys. Rev. B {\bf 61}, 11285 (2000);
D. Duffy, P. J. Hirschfeld, and D. J. Scalapino, {\bf 64}, 224522 (2001).

\bibitem{Matsukawa96}
M. Matsukawa, T. Mizukoshi, K. Noto, and Y. Shiohara, Phys. Rev. B {\bf 53}, 6034 (1996).

\bibitem{Hosseini98}
A. Hosseini, S. Kamal, D. A. Bonn, R. Liang, and W. N. Hardy, Phys. Rev. Lett.
{\bf 81}, 1298 (1998).

\bibitem{WuCarbotte}
W. C. Wu and J. P. Carbotte, Phys. Rev. B {\bf 57}, 5614 (1998).

\bibitem{ssvr}
B. S. Shastry and B. I. Shraiman, Phys. Rev. Lett. {\bf 65}, 1068 (1990);
Int. Journ. Mod. Phys. B {\bf 5}, 365 (1991);
A. Virosztek and J. Ruvalds, Phys. Rev. B {\bf 45}, 347 (1992).

\bibitem{tpdjkf1}
J. K. Freericks and T. P. Devereaux, Phys. Rev. B {\bf 64}, 125110 (2001).

\bibitem{tpdjkf2}
J. K. Freericks, T. P. Devereaux, and R. Bulla, Phys. Rev. B
{\bf 64}, 233114 (2001).

\bibitem{Manske}
T. Dahm, D. Manske, and L. Tewordt, Phys. Rev. B {\bf 59}, 14740 (1999).

\bibitem{NAFL}
T. P. Devereaux and A. P. Kampf, Phys. Rev. B {\bf 59}, 6411 (1999).

\bibitem{resonant}
More generally, in off-resonance conditions the vertices can be classified by
symmetry in terms of BZ or FS harmonics.

\bibitem{DMFT}
See, e.g., A. Georges, G. Kotliar, W. Krauth, and M. J. Rozenberg
Rev. Mod. Phys. {\bf 68}, 13-125 (1996).

\bibitem{ARPES2}
S. V. Borisenko, A. A. Kordyuk, S. Legner, C. D\"urr, M. Knupfer, M. S. Golden,
J. Fink, K. Nenkov, D. Eckert, G. Yang, S. Abell, H. Berger, L. Forr\'o,
B. Liang, A. Maljuk, C. T. Lin, and B. Keimer, Phys. Rev. B {\bf 64}, 094513 (2001);
P. V. Bogdanov, A. Lanzara, X. J. Zhou, S. A. Kellar, D. L. Feng, E. D. Lu,
H. Eisaki, J.-I. Shimoyama, K. Kishio, Z. Hussain, and Z. X. Shen; {\it ibid},
180505 (2001); A. A. Kordyuk, S. V. Borisenko, T. K. Kim, K. A. Nenkov, M. Knupfer,
J. Fink, M. S. Golden, H. Berger, and R. Follath
Phys. Rev. Lett. {\bf 89}, 077003 (2002).

\bibitem{Chubukov}
A. J. Millis, H. Monien, and D. Pines, Phys. Rev. B {\bf 42}, 167 (1990);
S. Sachdev and J. Ye, Phys. Rev. Lett. {\bf 69}, 2411 (1992);
R. Haslinger, A. V. Chubukov, and A. Abanov, Phys. Rev. B {\bf 63}, 20503 (2001).

\bibitem{DiCastro}
C. Castellani, C. Di Castro, and M. Grilli, Phys. Rev. Lett. {\bf 75}, 4650 (1995);
S. Caprara, M. Sulpizi, A. Bianconi, C. Di Castro, and M. Grilli, Phys. Rev. B
{\bf 59}, 14980 (1999).

\bibitem{Pomeranchuk1}
W. Metzner, D. Rohe, S. Andergassen, cond-mat/0303154.

\bibitem{extended}
D. Poilblanc, D. J. Scalapino, and W. Hanke, Phys. Rev. Lett. {\bf 72}, 884 (1994);
A. P. Kampf and T. P. Devereaux, Phys. Rev. B {\bf 56}, 2360 (1997).

\bibitem{Iyengar}
A. Iyengar, J. Stajic, Y.-J. Kao, and K. Levin, Phys. Rev. Lett. {\bf 90}, 187003 (2003).

\bibitem{nem}
S. A. Kivelson, E. Fradkin, and V. J. Emery, Nature {\bf 39}
550 (1998);
E. W. Carlson, V. J. Emery, S. A. Kivelson, and D. Orgad, cond-mat/0206217.

\bibitem{Wen}
X. G. Wen, Int. J. Mod. Phys. B {\bf 4}, 239 (1990);
P. A. Lee and N. Nagaosa, Phys. Rev. B {\bf 46}, 5621 (1992);
X.-G. Wen and P. A. Lee, Phys. Rev. Lett. {\bf 80}, 2193 (1998);
T. Senthil and M. P. A. Fisher, Phys. Rev. Lett. {\bf 86}, 292 (2000); {\it ibid.}
Phys. Rev. B {\bf 62}, 7850 (2000).

\bibitem{patch}
N. Furukawa, T. M. Rice, and M. Salmhofer, Phys. Rev. Lett. {\bf 81}, 3195 (1998);
D. Zanchi and H.J. Schulz, Phys. Rev. B {\bf 54}, 9509 (1996); ibid. {\bf 61}, 13609 (2000).
C. J. Halboth and W. Metzner, Phys. Rev. B {\bf 61}, 7364 (2000);
C. Honerkamp, M. Salmhofer, N. Furukawa, and T.M. Rice, Phys. Rev. B {\bf 63}, 035109
(2001); A. Katanin and A. P. Kampf, cond-mat/0304189.

\bibitem{Pomeranchuk2}
C. J. Halboth and W. Metzner, Phys. Rev. Lett. {\bf 85}, 5162 (2000);
B. Valenzuela and M. A. H. Vozmediano, Phys. Rev. B {\bf 63}, 153103 (2001);
C. Honerkamp, M. Salmhofer, and T. M. Rice, Eur. Phys. J. B {\bf 27}, 127 (2002);
A. Neumayr and W. Metzner, Phys. Rev. B {\bf 67}, 35112 (2003).

\bibitem{ARPES3}
T. Valla, A. V. Feodorov, P. D. Johnson, Q. Li, G. D. Gu, and N. Koshizuka, Phys. Rev. Lett. {\bf 85},
828 (2000).

\bibitem{MillisDrew}
A. J. Millis and D. Drew, cond-mat/0303018.

\bibitem{Hackl2002}
F. Venturini, Q.-M. Zhang, R. Hackl, A. Lucarelli, S. Lupi, M. Ortolani, P. Calvani,
N. Kikugawa, and T. Fujita, Phys. Rev. B {\bf 66}, 060502 (2002); private communication.

\bibitem{Ino}
A. Ino, C. Kim, M. Nakamura,T. Yoshida, T. Mizokawa, Z.-X. Shen, A. Fujimori, T. Kakeshita, H. Eisaki,
and S. Uchida, Phys. Rev. B {\bf 62}, 4137 (2000).

\bibitem{Lucarelli}
A. Lucarelli, S. Lupi, M. Ortolani, P. Calvani, P. Maselli, M. Caprizzi, P. Giura, H. Eisaki, N.
Kikugawa, T. Fujita, M. Fujita, and K. Yamada, Phys. Rev. Lett. {\bf 90}, 37002 (2003).

\bibitem{Lupi}
S. Lupi {\it et al.}, Phys. Rev. B {\bf 62}, 12418 (2000).

\bibitem{Munzar}
D. Munzar and M. Cardona, Phys. Rev. Lett. {\bf 90}, 077001 (2003).

\bibitem{DurstLee}
A. C. Durst and P. A. Lee, Phys. Rev. B {\bf 62}, 1270 (2000).

\bibitem{tpdapk2000}
T. P. Devereaux and A. P. Kampf, Phys. Rev. B {\bf 61}, 1490 (2000).

\bibitem{Chiao2000}
M. Chiao, R. W. Hill, C. Lupien, L. Taillefer, P. Lambert, R. Gagnon, and
P. Fournier, Phys. Rev. B {\bf 62}, 3554 (2000).

\bibitem{dvz}
T. P. Devereaux and D. Einzel, {\bf 51}, 16336 (1995);
T. P. Devereaux, A. Virosztek and A. Zawadowski, Phys. Rev. B {\bf 54}, 12523 (1996).

\bibitem{Tmatrix}
This
treatment does not consider the interplay of disorder and interactions
(and particular misses out
on the physics of the Anderson-Mott transition) and can only be considered
this way in the limit of weak scattering. Efforts to include both disorder
and interactions equally in a T-matrix approach have been put forward, most
recently in Ref. \cite{tpdapk2000}.

\bibitem{Graf96}
M. J. Graf, S.-K. Yip, J. A. Sauls, and D. Rainer, Phys. Rev. B {\bf 53},
15147 (1996).

\bibitem{Graf95}
M. J. Graf, M. Palumbo, D. Rainer, and J. A. Sauls, Phys. Rev. B {\bf 52},
10588 (1995).

\bibitem{Latyshev99}
Y. I. Latyshev, T. Yamashita, L. N. Bulaevskii, M. J. Graf, A. V.
Balatsky, and M. P. Maley, Phys. Rev. Lett. {\bf 82}, 5345 (1999).

\bibitem{Berlinsky}
A. J. Berlinsky, D. A. Bonn, R. Harris and C. Kallin, Phys. Rev. B {\bf 61}, 9088 (2000).

\bibitem{Atkinson}
W. A. Atkinson and P. J. Hirschfeld, Phys. Rev. Lett. {\bf 88}, 187003 (2002).

\bibitem{Quinlan94}
S. Quinlan, D. J. Scalapino, and N. Bulut, Phys. Rev. B {\bf 49}, 1470 (1994).

\bibitem{Quinlan97}
S. M. Quinlan, P. J. Hirschfeld, and D. J. Scalapino, Phys. Rev. B
{\bf 53}, 8575 (1996).

\bibitem{shastry}
B. S. Shastry, Phys. Rev. Lett. {\bf 63}, 1288 (1989); D. C. Mattis,
Phys. Rev. Lett. {\bf 74}, 3676 (1995).

\bibitem{jefferson}
J. H. Jefferson, H. Eskes, and L. F. Feiner, Phys. Rev. B {\bf 45}, 7959 (1992).

\end{thebibliography}
\end{document}